\documentclass[aps,prb,twocolumn,citeautoscript,superscriptaddress,pdflatex=1]{revtex4-1}
\usepackage{amsmath,amssymb,mathrsfs,bm,feynmf,setspace}
\usepackage{graphicx}
\usepackage[tight]{subfigure}
\usepackage{color}
\usepackage[papersize={8.5in,11in}]{geometry}
\usepackage[colorlinks=true]{hyperref}
\hypersetup{
    bookmarks=true,         % show bookmarks bar?
    unicode=false,          % non-Latin characters
    pdftoolbar=true,        % show Acrobat
    pdfmenubar=true,        % show Acrobat
    pdffitwindow=false,     % window fit to page when opened
    pdfstartview={FitH},    % fits the width of the page to the window
    pdftitle={My title},    % title
    pdfauthor={Author},     % author
    pdfsubject={Subject},   % subject of the document
    pdfcreator={Creator},   % creator of the document
    pdfproducer={Producer}, % producer of the document
    pdfkeywords={keyword1} {key2} {key3}, % list of keywords
    pdfnewwindow=true,      % links in new window
    colorlinks=true,       % false: boxed links; true: colored links
    linkcolor=red,          % color of internal links (change box color with linkbordercolor)
    citecolor=blue,        % color of links to bibliography
    filecolor=magenta,      % color of file links
    urlcolor=cyan           % color of external links
}
\definecolor{palatinatepurple}{rgb}{0.41, 0.16, 0.38}

\definecolor{uglybrown}{rgb}{0.8,  0.7,  0.5}

\numberwithin{equation}{section}

\newcommand{\beq}{\begin{equation}}
\newcommand{\eeq}{\end{equation}}
\def\bea{\begin{eqnarray}}
\def\eea{\end{eqnarray}}

\newcommand{\tr}{\text{tr}}

\def\({\left(}
\def\){\right)}

\def\CI{{\cal I}}

\def\CN{{\cal N}}
\def\CO{{\cal O}}%AEL

\def\CR{{\cal R}}

\begin{document}

\title{Entanglement structure of non-equilibrium steady states}
  \author{Raghu Mahajan}
 \affiliation{Department of Physics, Stanford University, Stanford CA 94305, USA }
 \author{C. Daniel Freeman}
 \affiliation{Department of Physics, University of California at Berkeley, Berkeley CA 94720, USA}
 \author{Sam Mumford}
 \affiliation{Department of Physics, Stanford University, Stanford CA 94305, USA }
  \author{Norm Tubman}
 \affiliation{Department of Chemistry,  University of California at Berkeley, Berkeley CA 94720, USA}
 \author{Brian Swingle}
  \email{bswingle@stanford.edu}
 \affiliation{Department of Physics, Stanford University, Stanford CA 94305, USA }
 \date{\today}

\begin{abstract}

We study the problem of calculating transport properties of interacting quantum systems, specifically electrical and thermal conductivities, by computing the non-equilibrium steady state (NESS) of the system biased by contacts. Our approach is based on the structure of entanglement in the NESS. With reasonable physical assumptions, we show that a NESS close to local equilibrium is lightly entangled and can be represented via a computationally efficient tensor network. We further argue that the NESS may be found by dynamically evolving the system within a manifold of appropriate low entanglement states. A physically realistic law of dynamical evolution is Markovian open system dynamics, or the Lindblad equation.
We explore this approach in a well-studied free fermion model where comparisons with the literature are possible. We study both electrical and thermal currents with and without disorder, and compute entropic quantities such as mutual information and conditional mutual information. We conclude with a discussion of the prospects of this approach for the challenging problem of transport in strongly interacting systems, especially those with disorder.

\end{abstract}
\maketitle
\newpage
\section{Introduction and overview}

Measurements of electrical and thermal currents access a rich store of information about a quantum many-body system\cite{Datta1997,Blandin1976,Spohn1980,Leggett1987,Han2013}: they inform us about equilibrium thermodynamics, sources of dissipation, scattering timescales, and the number and nature of conservation laws. Treating all of this physics appropriately makes theoretical calculation of electrical and thermal currents challenging. In many cases, there is a general linear response formalism, but the calculation of the necessary real time correlation functions at finite temperature and in the presence of interactions is difficult. This is to say nothing of systems further out of equilibrium.

For systems with a quasi-particle description, a kinetic or Boltzmann equation approach may suffice, but many strongly interacting systems have no quasi-particles. Hydrodynamics may provide a good description of the long wavelength physics, but the calculation of the transport coefficients is still difficult in general\cite{Spohn1980}. Numerical approaches are challenging for a variety of reasons, for example, the real time nature of the physics is in tension with imaginary time approaches. Direct time evolution of the many-body state is computationally infeasible in all but a select set of physical problems\cite{Vidal2004}.

Our approach is to directly model the non-equilibrium steady state (NESS) of the system using \emph{real-time}, \emph{open-system} evolution. Despite many approaches for treating non-equilibrium steady states~\cite{Datta1997,Lothar2008,Pletyukhov2010,Egger2000,Blandin1976,Molmer1993,Dalibard1992,Gardiner1992,Prosen2008}, development of new methods remains an active field of research. One of the most widely used methods is the non-equilibrium Green's function approach, which can be used quite generally, but it is essentially a perturbative scheme, and is not expected to be accurate for strongly interacting systems~\cite{Haug1996}.

Our specific interest is transport in strongly interacting systems. Away from the free particle limit, exact diagonalization is feasible only for small systems. There are many sorts of quantum Monte Carlo techniques that can treat interacting systems, but they are generally only suitable for imaginary time propagation. There has been some progress developing real time propagation with a quantum Monte Carlo framework~\cite{Lothar2008}. However sampling and sign problem issues~\cite{Tubman2011} have to be considered for each Hamiltonian individually.
In some strongly interacting systems with weakly broken translation invariance, a kinematic approach to transport using the so called ``memory-matrix" method is possible
\cite{Hartnoll:2012rj, Mahajan:2013cja, Hartnoll:2014gba, Lucas:2015pxa}.
There are also special systems in (2+1)-dimensions 
involving strong Chern-Simons type interactions in which 
it is possible to calculate the transport-coefficients exactly in the planar limit 
\cite{Giombi:2011kc, GurAri:2012is, Aharony:2012nh, Gur-Ari:2016xff}.
However, one would like to access a wider variety of systems.

The key physical idea of this work is that, in many contexts, NESS are lightly entangled~\cite{Prosen2015,Blythe2007,Alicki2002}, at least if they are in local thermal equilibrium\cite{Cui2015}. One does not expect substantially more entanglement than in the ground state and in many cases, e.g. at finite temperature or in contact with leads, there may be substantially less entanglement than in the ground state\cite{Verstraete2004}. Specifically, we would expect NESS to have at most area law entanglement at any non-zero temperature: while the von Neumann entropy may be extensive in system size, this extensive entropy is thermodynamic in origin, and the actual entanglement present obeys an area law. Furthermore, these intuitions strongly suggests that NESS should have a computationally efficient representation using so-called tensor networks.

The entanglement based approach to certain many-body states based on matrix product representations (a special kind of tensor network) is well established in one spatial dimension~\cite{White1992,Schollwock2005,Schollwock2011,Verstraete2004,Hartmann2009}. There is numerical evidence~\cite{Cui2015,Mascarenhas2015,Jin2013,Weimer2015,Cai2013,Kliesch2014,Prosen2010} and evidence from integrable systems~\cite{Karevski2013,Clark2010,Prosen2009,Prosen2012,Prosen2015,Prosen2011,Prosen2011-1,Prosen2014,Prosen2012-1,Popkov2015} that NESS can be described using matrix product states (MPS). Both unitary real-time evolution and open system evolution can also be carried out in the context of MPS~\cite{Vidal2004,Zwolak2004,White2004,Daley2004}. There are also a few discussions of open system dynamics and NESS in higher dimensions~\cite{Yan2011,Stoudenmire2012}.

To go beyond the intuitive picture sketched above and the well-established one-dimensional approaches, we use tools from quantum information theory, specifically the theory of approximate quantum Markov chains \cite{qmarkov,fr1,fr2,2015RSPSA.47150338W,2015arXiv150907127J}, to establish the claim that near-equilibrium NESS are lightly entangled in many contexts. The output of this analysis is an existence proof of a tensor network representation for the state given certain entropic assumptions. Once we know the NESS is lightly entangled and has an efficient tensor network representation, how do we find the NESS? There are a number of approaches to this problem including variational and dynamical; here we focus on the latter possibility, namely determining the NESS by dynamically evolving the system to its steady state.

To model the dynamics of the system coupled to an environment we use Markovian dissipative dynamics. The density matrix $\rho$ of the system of interest is assumed to evolve according to a Lindblad equation
\beq \label{lindblad}
\partial_t \rho = [-i H, \rho] + \sum_j L_j \rho L_j^\dagger - \frac{1}{2} \sum_j \{ L_j^\dagger L_j, \rho\}.
\eeq
Here $H$ is the Hamiltonian describing the evolution of the system and the $L_j$ are so-called jump operators which describe the dissipative effects of the environment. Several tools have been developed to directly solve Lindblad master equations~\cite{Schack1997,Tan1999,Vukics2007,Johansson2012}. For small systems, direct diagonalization and integral techniques are the most efficient methods to use. In a benchmark study of Heisenberg spin chains~\cite{Johansson2012}, once a critical system size is reached, roughly about 10 sites, wave function Monte Carlo becomes a more efficient method. However, as discussed above, even Monte Carlo is not always suitable.

As already mentioned, the use of tensor network methods to treat NESS looks promising. There have been significant efforts to develop 1d methods known as matrix product operators (MPOs) for explicitly time evolving the Lindblad equation. The use of MPO for solving NESS applications had been originally proposed in the context of bulk dissipation~\cite{Zwolak2004}. Since then it has been applied to a number of different models. For instance, when assuming an MPO ansatz for the solution of a NESS, various properties can be proved for both spin systems~\cite{Prosen2011-1,Ilievski2014}, and for the Hubbard model~\cite{Popkov2015}. For spin systems, numerical results have been used to study integrable and non-integrable models for system sizes on the order of hundreds of sites such as in the Heisenberg XXZ Hamiltonian~\cite{Prosen2009} and the Heisenberg XY~\cite{Prosen2008-1,Prosen2011}. There have been some studies of NESS for the Hubbard model, such as looking at transport at infinite temperature. These simulations have also been performed with system sizes on the order of hundreds of sites~\cite{Prosen2012}. This is an active field: one needs to improve accuracy, use larger system sizes and reach lower temperatures.

We now present the results of this paper in more detail. Our first main result is a systematic study of open system transport in a simple one-dimensional free fermion model. This serves as a benchmark for our approach as it allows us to model the bath in a simple way and probe the temperature range accessible to our methods. Of course this is a well-studied class of models, although we are not aware of the precise setup we consider being studied before. We study both electrical and thermal currents with and without disorder and compare to known results obtained from the scattering theory approach \cite{5392683,quantumtransportbook}. To support our general information theoretic arguments, we also calculate entropic measures of correlation in the free fermion NESS we study. These measures include the mutual information, which we use to establish the decay of correlations, and the conditional mutual information, which we use to establish that the free fermion NESS is an approximate quantum Markov chain \cite{qmarkov}.

Our second main result concerns general local Hamiltonians and establishes that if a NESS has the property of local thermal equilibrium, then an efficient tensor network representation of this NESS exists. Our argument is based on the entropic structure arising from local thermal equilibrium and uses the theory of quantum Markov chains \cite{qmarkov,fr1,fr2,2015RSPSA.47150338W,2015arXiv150907127J} building on a recent construction  \cite{2016arXiv160705753S}. The necessary entropy structure is physically reasonable and we observe that both free fermion NESS (from our numerics) and holographic NESS (using AdS/CFT duality \cite{1999IJTP...38.1113M,2006PhRvL..96r1602R,2008JHEP...02..045B}) satisfy it. We also argue for the existence of a further renormalization group (RG) structure in the NESS tensor network, but we do not rigorously show its existence.

In the discussion we give an argument that the tensor network approach should open up new vistas in the calculation of transport properties of strongly interacting quantum many-body systems. Our argument is supported by and builds on the existing successes of the tensor network approach to NESS in 1d open systems. In future work, we will build on these results by using the machinery of tensor networks to study open-system dynamics with strong interactions and by investigating variational methods to find steady states. The appendices contain some technical details and reviews of certain topics like the scattering approach to 1d transport.

\section{Entanglement and currents in a free fermion model}
\label{ffmodel}

In this section we study a free fermion model of transport. The physical setup consists of some number of fermion modes evolving according to a Lindblad equation where the Hamiltonian $H$ is quadratic in the fermion modes and the jump operators are linear in the fermion modes. Such a many-body Lindblad equation is reducible to a dynamical equation for the single particle Green's function. This makes the problem numerically tractable.  Based on this framework we present the form of the current in an analytically solvable two-site model and numerical calculations of currents in a general multi-site model.

In the multi-site model, we discuss several features including the temperature dependence of conductance and the effects of disorder. We compare our results to the Landauer scattering model of transport. Finally, as an illustration of the general information theoretic arguments in Section \ref{qimodel}, we study the entanglement structure of the free fermion NESS. In particular, we compute the mutual information, which is an upper bound on size of correlations in the state, and the conditional mutual information, which gives insight into the spatial structure of correlations. We show that the free fermion NESS is both short-range correlated and an approximate quantum Markov chain. The significance of these results will be expanded upon in Section \ref{qimodel}.

We use standard correlation matrix machinery, although the information theoretic results at the end of this section imply that tensor networks methods would also work. The use of the correlation matrix technology has the advantage of letting us explore the physics without the technical complications of tensor network methods.

\subsection{Setup}

The free fermion problem is specified by giving a list of fermion annihilation operators $\{c_\alpha \}$, a Hamiltonian $H$ and a set of jump operators $L_j$ built from the fermion operators. In order for the many-body fermion problem to reduce to a single particle problem the Hamiltonian must be quadratic in $c$ and $c^\dagger$ and the jump operators must be linear in $c$ and $c^\dagger$. Hence we may write the Hamiltonian as
\beq
H = c^\dagger h c \label{eq:lindblad0}
\eeq
and the jump operators as either
\begin{align}
L_j &= u^\dagger_j c, \quad \text{or}\label{eq:lindblad1}\\
L_j &= c^\dagger v_j \label{eq:lindblad2}.
\end{align}
In these equations, $c$, $u_j$ and $v_j$ are thought of as column vectors.
Assuming that the density matrix $\rho$ is Gaussian, of the form
$\rho \sim \exp(- c^\dagger k c)$, all physical quantities can be computed from the correlation matrix
\begin{align}
G_{\alpha,\beta} = \text{tr}(c_\alpha^\dagger c_\beta \rho).
\end{align}
Plugging the above forms of $H$, $L_j$ and $\rho$ into the Lindblad equation \eqref{lindblad}, it is possible to derive an equation of motion for the correlation matrix \cite{2013NJPh...15h5001B}, see Appendix \ref{fflindblad} for details. The result is
\begin{align}
\partial_t G = i [h^T,G] - \frac{1}{2}\sum_j \left\{u_j^* u_j^T,G\right\} + \frac{1}{2}\sum_j \left\{v_j^* v_j^T,1-G\right\}.
\label{eq:markovG}
\end{align}
Knowing $G$ is sufficient to compute densities, currents, and entropies. It also bears repeating that we could have approached this problem using tensor network methods, but to focus on the physics of the open system we use the simpler correlation matrix approach.

Let us introduce the general multi-site model we will study. Consider a one dimensional wire composed of three parts: the left contact $(L)$, the right contact $(R)$, and the central part of the wire $(C)$. The bare Hamiltonian for each piece of the wire is taken to be the same free fermion hopping model given by
\beq
H_0 = \sum_{x=1}^{N-1} - w (c_x^\dagger c_{x+1} + c_{x+1}^\dagger c_x)
\eeq
where $N = N_L, N_R, N_C$ for the left, right, or central parts, respectively.
The central part $C$ will be our system, and $L$ and $R$ will function as the contacts or baths. The three segments of wire are coupled by a hopping matrix element of strength $w'$ where they meet. Finally, we add jump operators in the $L$ and $R$ regions which drive those regions to the thermal state of their respective bare Hamiltonians, without considering the coupling $w'$. Note that we do not have Lindblad operators acting on the central part. The desired Lindblad operators are taken proportional to the energy eigenmodes of the uncoupled wires, with an overall coefficient $\gamma$. The relative size of the annihilation \eqref{eq:lindblad1} and creation \eqref{eq:lindblad2} terms is set by the Einstein relations. See Appendix \ref{ffthermallindblad} for more details.

Note that chemical potentials are not included in the Hamiltonian $H_0$. These potentials can be directly included in the jump operators; tuning the average chemical potential $(\mu_L + \mu_R)/2$ can set the overall density as desired. We may also generalize this model slightly by allowing more general system-bath couplings. For example, we may consider further neighbor hoppings between our system and the bath.

\subsection{Solvable two-site model}

Before presenting data for the general multi-site model, it instructive to solve the model with just two sites, labeled $L$ and $R$. The jump operators are given by the in-rates, $v_L = \sqrt{\gamma}$ and $v_R = \sqrt{\gamma}$, and the out-rates, $u_L = \sqrt{\gamma e^{\beta_L (\epsilon_L - \mu_L)}}$ and $u_L = \sqrt{\gamma e^{\beta_R (\epsilon_R - \mu_R)}}$. These rates lead to equilibrium distributions of the form
\beq
n_{L} = \frac{\gamma_{in}}{\gamma_{out}+\gamma_{in}} = \frac{1}{e^{\beta_{L}(\epsilon_{L}-\mu_{L})}+1},
\eeq
and likewise for $R$.
The above jump operators are already a special case since the base rate $\gamma$ is taken the same everywhere. Assume without loss of generality that $\epsilon_{L} = \epsilon_R = 0$, this may be compensated by $\mu_{L}, \mu_R$. Finally, for simplicity take $\beta_{L} = \beta_R =\beta$.

The Hamiltonian is taken to be $H = - w' c_L^\dagger c_R - w' c_R^\dagger c_L$. The quantities appearing in \eqref{eq:markovG} controlling the dissipative physics are
\begin{align}
\sum_j u_j^\star u_j^T &= \gamma \left(
                           \begin{array}{cc}
                             e^{-\beta \mu_L} & 0 \\
                             0 & e^{-\beta \mu_R} \\
                           \end{array}
                         \right),\\
\sum_j v_j^\star v_j^T &= \gamma \left(
                           \begin{array}{cc}
                             1 & 0 \\
                             0 & 1 \\
                           \end{array}
                         \right).
\end{align}
Let us define $r_0 + \delta r = e^{-\beta \mu_{L}}$ and
$r_0 - \delta r = e^{-\beta \mu_{R}}$.

The fixed point equation following from \eqref{eq:markovG} is easily solved (Appendix \ref{app:twosite}).
The current in units of energy is given by
\beq
\label{eq:twositecurrent}
I = - \frac{2 w'^2}{\gamma(1+r_0)} \frac{\delta r}{\frac{4 w'^2}{\gamma^2} + (1+r_0)^2 - \delta r^2}.
\eeq
Note that if $\delta r >0$, so that $n_L < n_R$, then $I <0$ and current flows from $R$ to $L$.

This equation for the current displays a lot of rich physics. Let us take the linear response limit of $\delta r \ll 1$. Then we can see the physics of the various rates very clearly.
If $\gamma \gg w'$ then the current goes like $I \sim w'^2 \delta r/\gamma$, indicating that the $\gamma$ parameter functions as the internal resistance of the contacts. The current drops to zero as $\gamma \rightarrow \infty$ because the left and right sites are then strongly forced to their equilibrium states.
If $\gamma \ll w'$, the current goes like $I \sim \gamma \delta r$ indicating that the bottleneck to charge flow is charge input from the Lindblad operators. These two limits also tell us that the current exhibits a maximum as a function of $\gamma$. In the multisite model, we will study the limit $\gamma \ll w'$, since otherwise the resistance is dominated by the contacts.

\subsection{Transport properties of the multi-site model}
\label{sec:ffdata}

In this subsection, we make a detailed numerical study of the multi-site model. Interesting features include recovery of many results from the scattering theory of 1d transport and diffusive transport in the presence of disorder. To solve this model, we rewrite the steady-state condition derived from
\eqref{eq:markovG} as a matrix-inversion problem. Then, we use a sparse linear solver to find the steady-state correlation matrix $G$. We note that other methods are feasible, for example, we have also studied a real time approach which directly integrates \eqref{eq:markovG} and which recovers the results reported below. A forthcoming paper \cite{Zanoci} will explore in more detail the physics of open system fermion models using a method developed by Prosen \cite{Prosen2008}.

We first study the electrical conductance $G$, working in the linear response regime. The conductance is defined as the ratio of the current $I$ to the voltage $V$, $G = I/V$; it is related in the usual way to the conductivity by a dimension dependent factor depending on the geometry of the system. The temperatures in the three sections of the system are all equal,
$\beta_L=\beta_R=\beta_C=2$. The average chemical potential is zero and a small bias $\mu_L - \mu_R$ is applied. We set $w=1$ and explore the parameter space of $w'$ and $\gamma$. The number of sites is $N_L=N_R=40$ and $N_C = 11$. The resulting $G = I/(\mu_L-\mu_R)$, with the current evaluated at the middle site, is plotted below.
\begin{figure}[!h]
\includegraphics[width=.49\textwidth]{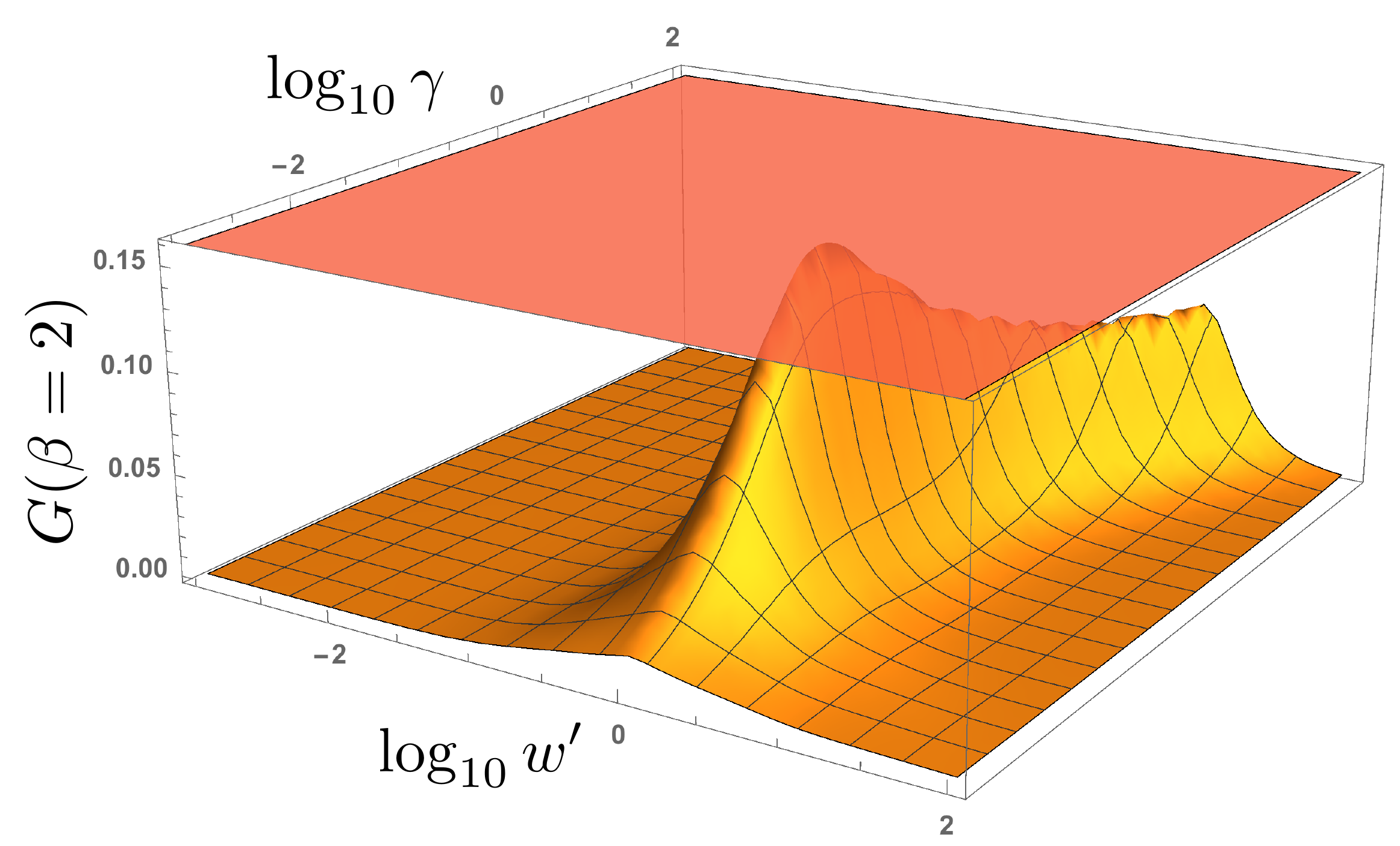}
\caption{The linear response conductance as a function of the Lindblad strength $\gamma$ and the lead-system coupling $w'$. The top horizontal surface is at $G = 1/(2\pi)$. This plot was made using $N_L=N_R=40$, $N_C=11$, $\beta_L=\beta_R=\beta_C=2$. The average chemical potential is zero.}
\label{fig:3dplot}
\end{figure}

We see a pronounced maximum, which sits roughly at $w'=1, \gamma=0.051$. The value of $G$ at this maximum is very close to the low-temperature scattering theory answer $G=1/(2\pi)$. From here on, we work with these values of $\gamma$ and $w'$.

\begin{figure}[!h]
\includegraphics[width=.49\textwidth]{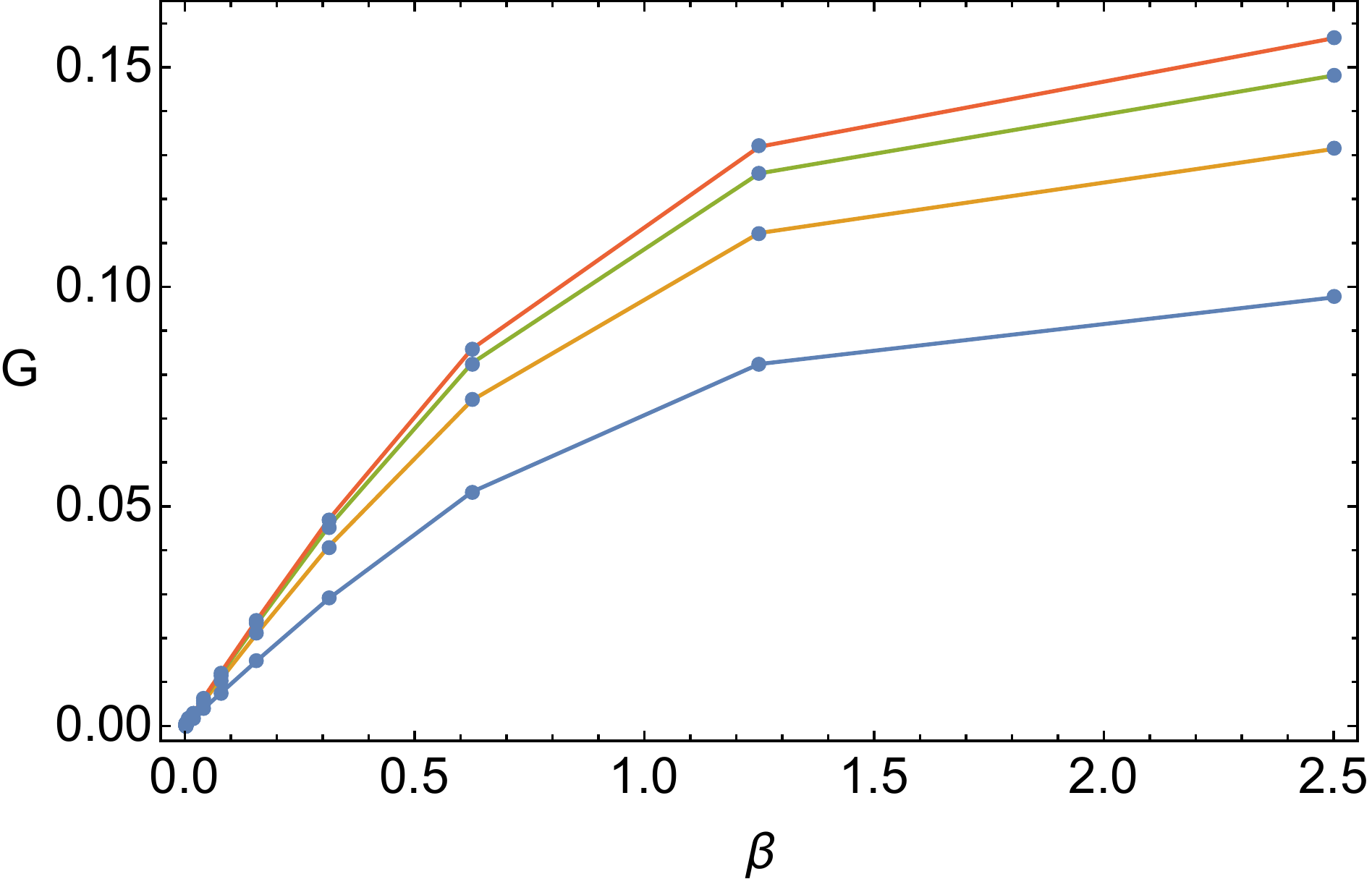}
\caption{The linear response conductance as a function of inverse temperature $\beta$. The different curves are for different system sizes, the system sizes being 31, 51, 71 and 91 sites as we go upwards. We have set $w'=1$ and $\gamma=0.051$.}
\label{fig:sigmavbeta}
\end{figure}

Next, we study the linear response conductance as a function of the temperature, depicted in Figure \ref{fig:sigmavbeta}.
We found that the multi-site systems recapitulated much of the qualitative physics encapsulated by the two-site model.
The linear response conductance scales linearly with $\beta$ at small $\beta$, and becomes independent of $\beta$ at large $\beta$.  This scaling matches the expected result from both the two-site model \eqref{eq:twositecurrent} as well as a simple argument from scattering theory, see Appendix \ref{sec:1dballistic}.

\begin{figure}[!h]
\includegraphics[width=.49\textwidth]{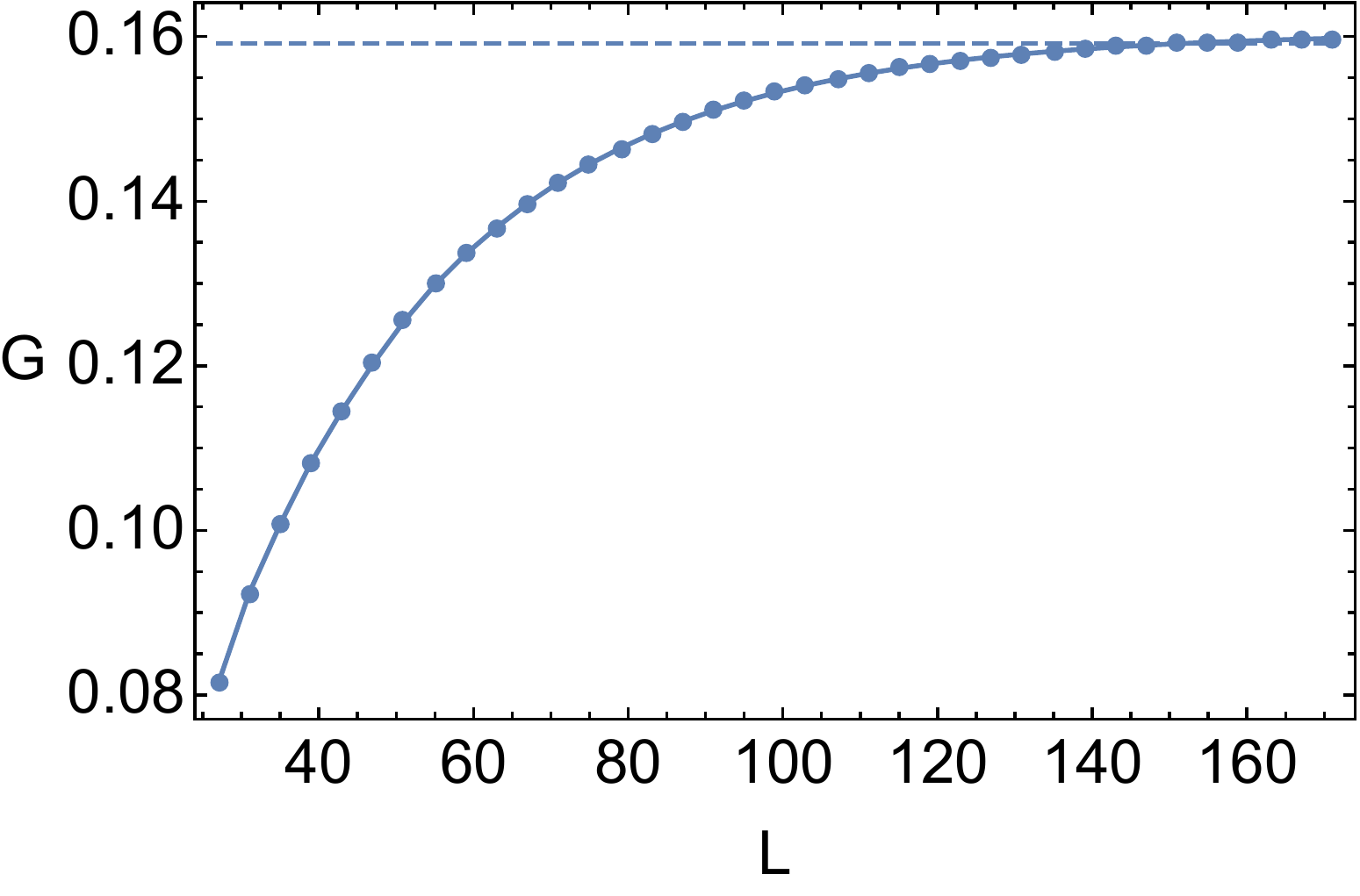}
\caption{The linear response conductance as a function of system size $L=N_L+N_R+N_C$. This plot is for a reasonably low temperature $\beta=3.0$. We see convergence of the results to the scattering theory answer in the thermodynamic limit.}
\label{fig:sigmavl}
\end{figure}

In Figure \ref{fig:sigmavl}, we show convergence of our results with system size.
The maximum system size we were able to reach was 171 sites.  Because the bath level spacing is related to the number of sites on the contacts, and because we sought results in the continuum limit, being constrained by a maximum number of sites effectively set a largest possible $\beta$. Once the temperature was smaller than this, the results were no longer representative of a continuum limit.
Regardless, 171 sites was enough to accurately converge the linear response conductance as depicted in Figure \ref{fig:sigmavl}.

\begin{figure}[!h]
\includegraphics[width=.49\textwidth]{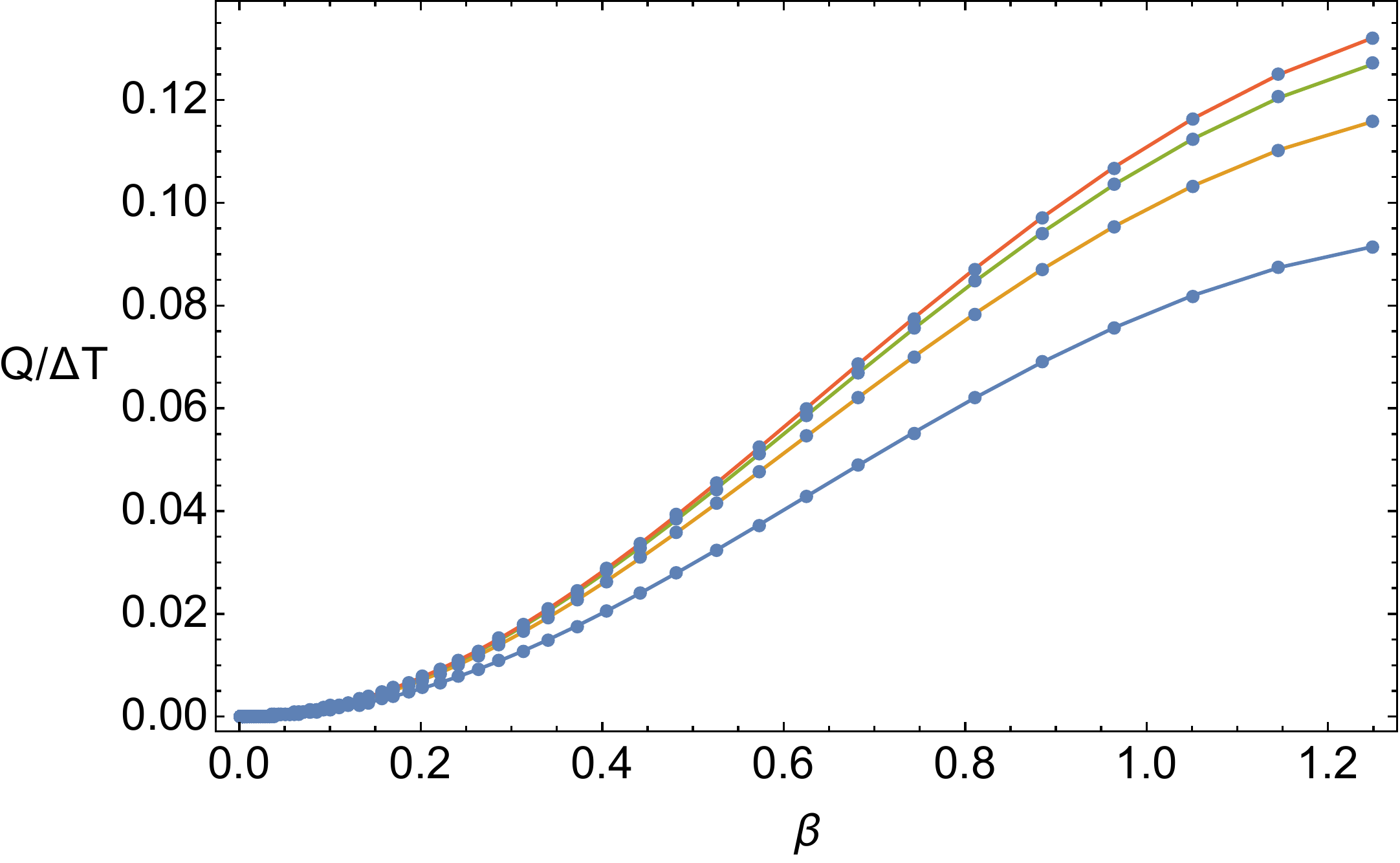}
\caption{The linear response thermal conductance as a function of inverse temperature. The different curves are for system sizes consisting of 31, 51, 71 and 91 sites as we go upwards.}
\label{fig:kappavsmallbeta}
\end{figure}

Next we set all chemical potentials to zero, and introduce a small temperature bias
$\Delta T = T_L-T_R$. This enables us to compute the thermal current $Q$ and study the linear response thermal conductance $Q/\Delta T$ as a function of temperature.
The results are shown in Figure \ref{fig:kappavsmallbeta}. The quadratic scaling at low
$\beta$ is again in accordance with the expected results from scattering theory, see Appendix \ref{sec:1dballistic}.

\begin{figure}[!h]
\includegraphics[width=.49\textwidth]{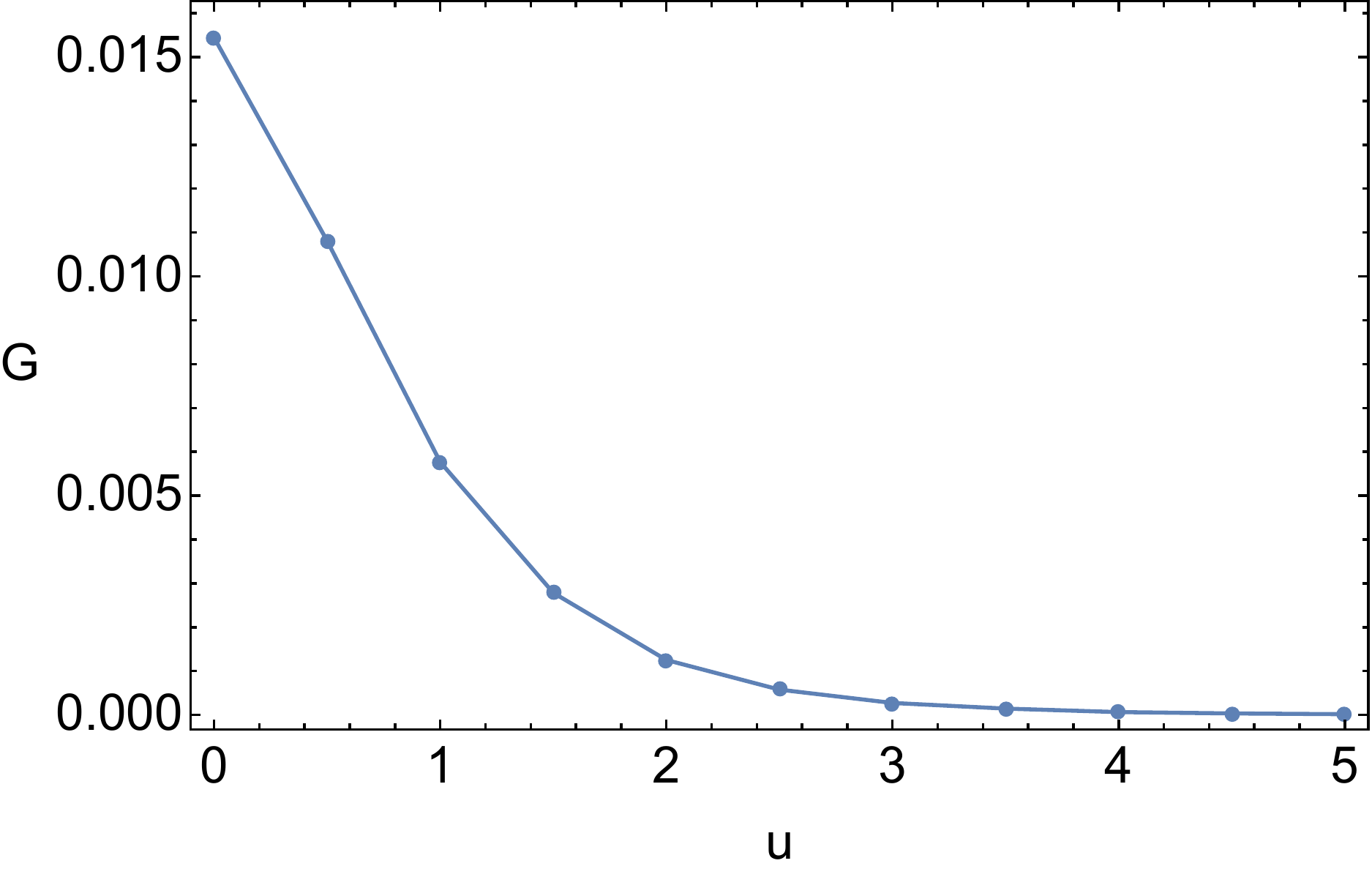}
\caption{The linear response conductance as a function of disorder strength $u$.
The chemical potential at each site is picked from a flat distribution in the interval $[-u,u]$. As expected, the conductance drops to zero as the disorder strength increases.}
\label{fig:sigmavdisorder}
\end{figure}

\begin{figure}[!h]
\includegraphics[width=.49\textwidth]{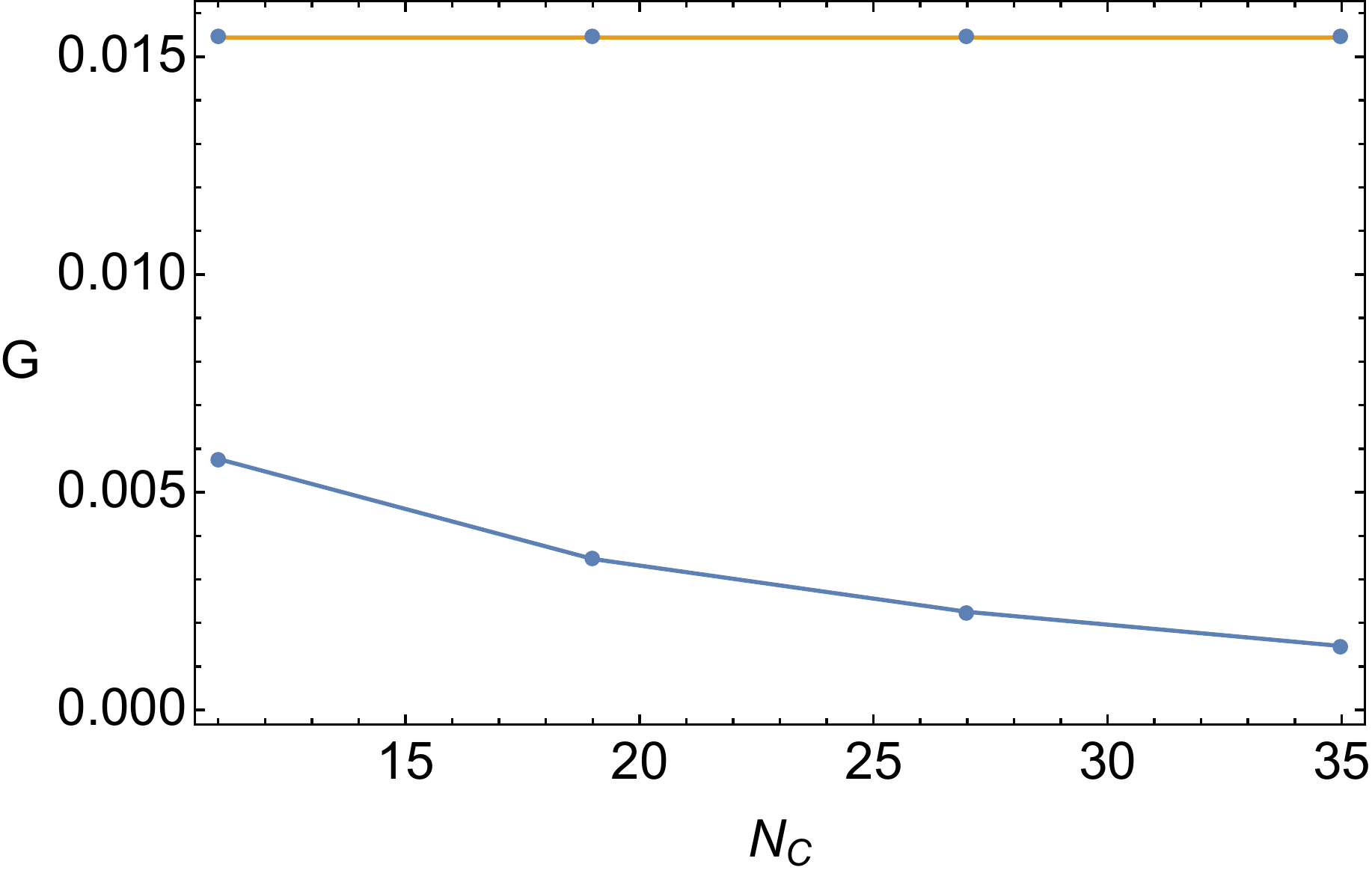}
\caption{The linear response conductance as a function of system size in the clean system (upper curve) and in the presence of disorder (lower curve). We see that in the presence of disorder, the conductance vanishes in the thermodynamic limit, as expected from localization.}
\label{fig:sigmavldisorder}
\end{figure}

Having explored the clean system, we now study the disordered case. We disorder the system by adding a random chemical potential at each site. The randomness distribution is taken to be flat, with values between $-u$ and $u$.
Linear response data for conductance as a function of $u$ is shown in Figure \ref{fig:sigmavdisorder}. We see that increasing the disorder leads to a drop in the conductance, as expected. Further, in Figure \ref{fig:sigmavldisorder}, we compare the system-size dependence of a clean and disordered wire. In the clean system, conductance is independent of system size indicating the ballistic nature of the transport, whereas in the disordered system, conductance goes to zero in the thermodynamic limit indicating the diffusive nature of the transport.

\subsection{Mutual Information}

In this subsection, we study the entanglement structure of the free fermion NESS. We study the mutual information, which upper bounds all classical and quantum correlations, and the conditional mutual information, which gives information about the spatial structure of entanglement. The observations here provide context for the more general results in Section \ref{qimodel}. We provide some context for these results in Appendix \ref{app:entropyboosted}.

\begin{figure}[!h]
\includegraphics[width=.49\textwidth]{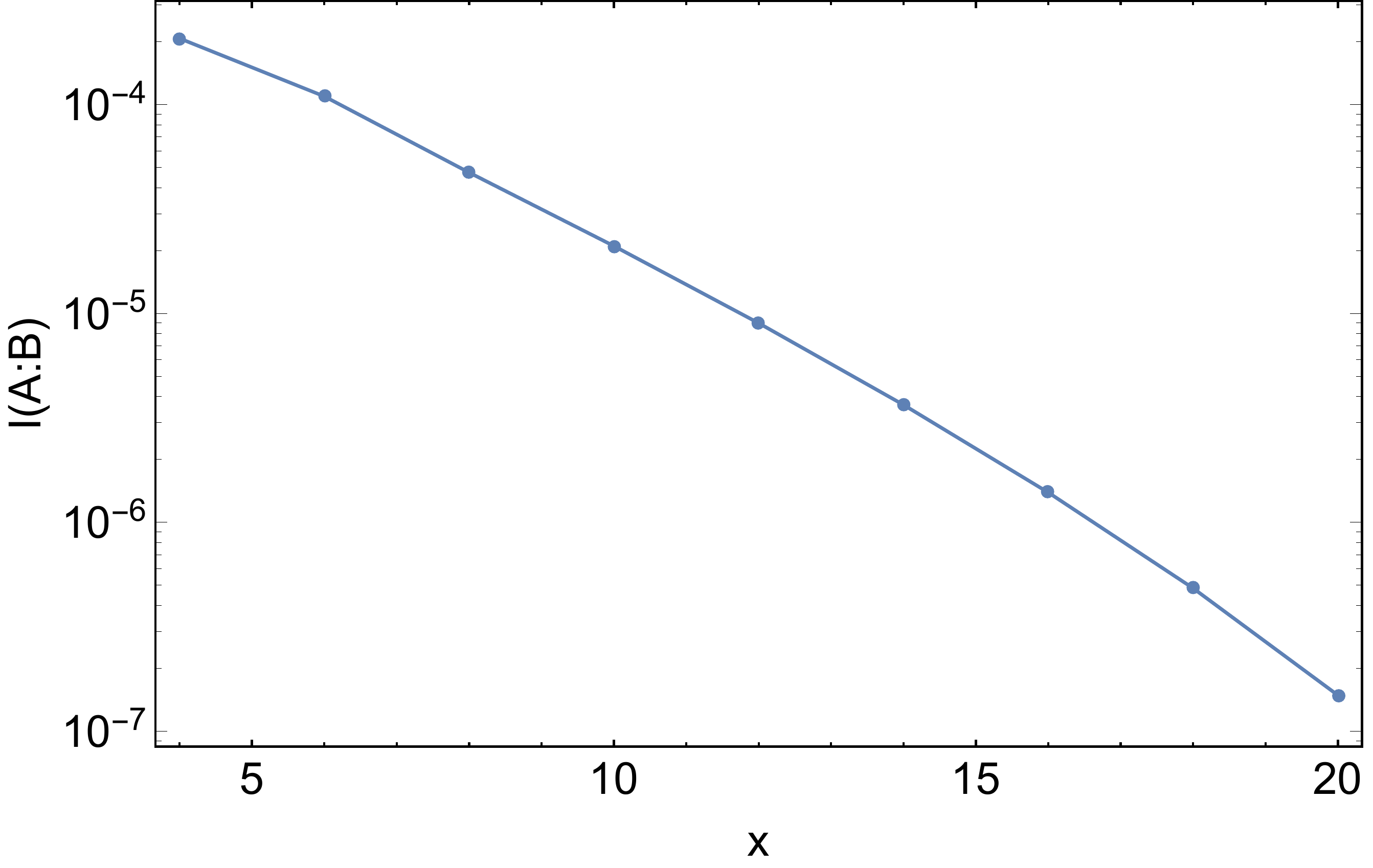}
\caption{The mutual information $I(A:B)$ as a function of the separation $x$ between two small regions $A$ and $B$ in the system. We took the subsystems $A$ and $B$ to have five sites each and $\beta=1$. This plot clearly shows that the mutual information decays quickly as the separation between the systems increases.}
\label{fig:mutualinformation}
\end{figure}

The mutual information of two subsystems $A$ and $B$ is defined as
\begin{align}
\mathcal{I}(A:B) := S(A)+S(B)-S(AB),
\end{align}
where $S$ denotes the von Neumann entropy of a region. We computed the mutual information between two small intervals as a function of the separation between the intervals. The results, shown in Figure \ref{fig:mutualinformation}, establish the fact that there is little correlation between far-separated regions. The free fermion NESS thus has only short-range entanglement.

\begin{figure}[!h]
\includegraphics[width=.49\textwidth]{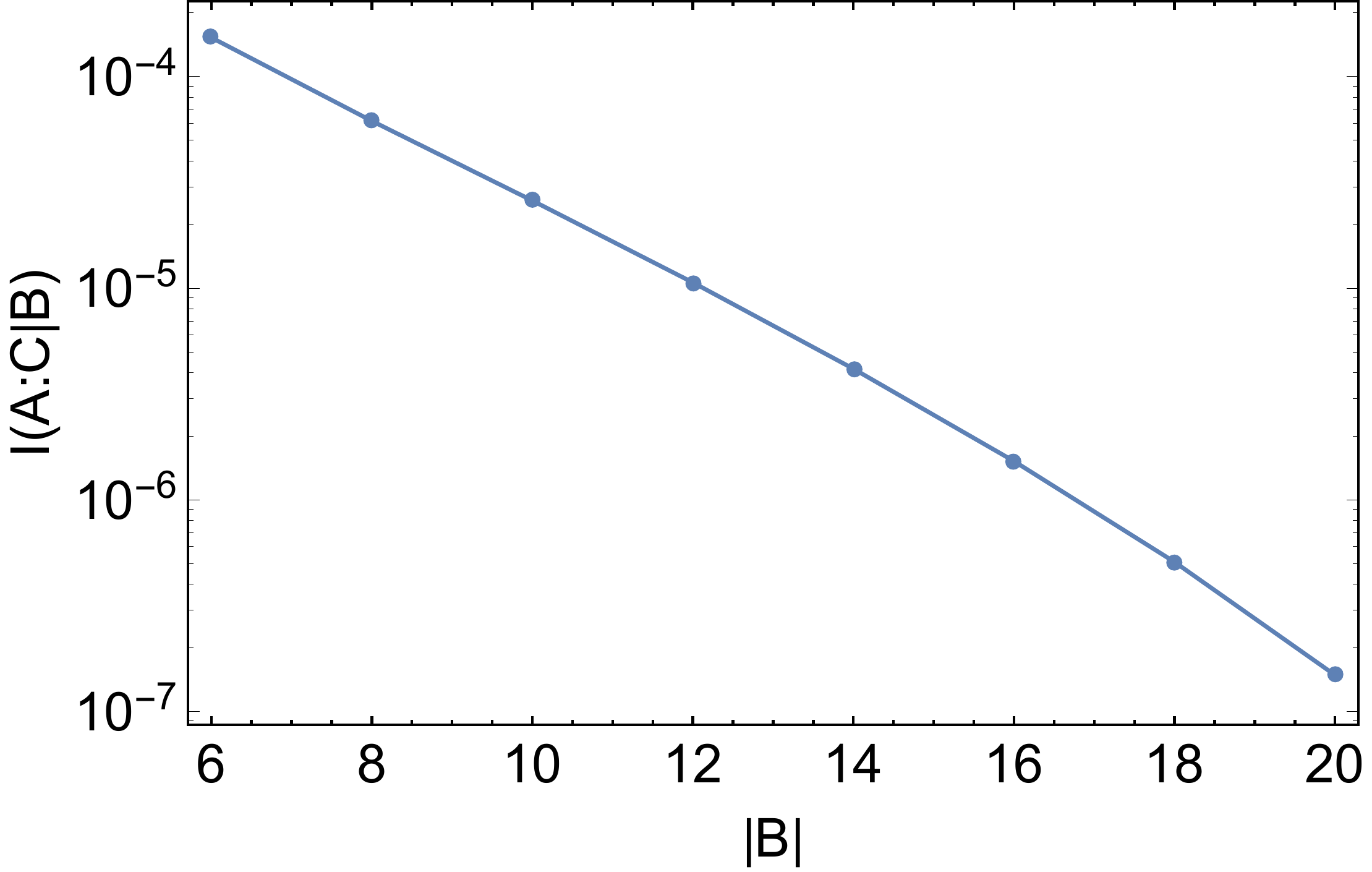}
\caption{The conditional mutual information $I(A:C|B)$ between adjacent regions $A$, $B$ and $C$ as a function of the size $\vert B \vert$ of the middle region. The regions $A$ and $C$ have fixed size of 5 sites each. This plot clearly shows that the conditional mutual information decays quickly as the separation increases.}
\label{fig:conditionalmutualinformation}
\end{figure}

The conditional mutual information between three subsystems $A$, $B$ and $C$ is defined as
\begin{align}
\CI(A:C|B) &:= \mathcal{I}(A:BC) - \mathcal{I}(A:B) \nonumber \\
&= S(AB) + S(BC) - S(B) - S(ABC)
\label{eq:cmidef}
\end{align}
This quantity is always positive due to the strong subadditivity of the von Neumann entropy. We computed this quantity in the free fermion NESS using three adjacent subsystems $A$, $B$ and $C$. The regions at the ends, $A$ and $C$, are taken to have five sites each, and we vary the size of system $B$. The results are shown in Figure
\ref{fig:conditionalmutualinformation}. We see that the conditional mutual information decays exponentially with the size of $B$. Let us now turn to the implications of this result, drawing from the theory of quantum Markov chains.

\section{NESS entanglement in general thermalizing many-body systems}
\label{qimodel}

Above we established that NESS were lightly entangled states in the limit of zero interaction. In this section we consider interacting systems and establish that the same structure of quantum correlations is obtained if the system reaches local thermal equilibrium.

Technically, we assume a particular form of the entropy of subsystems and show that this form implies that the NESS is an approximate quantum Markov chain. Using recent results on the structure of approximate quantum Markov chains, we establish that the NESS has an efficient tensor network representation. To give a non-trivial example of this construction, we briefly discuss NESS arising within the AdS/CFT correspondence.

We also comment on a possible refinement which incorporates an additional renormalization group (RG) structure into the tensor network down to microscopic scales. Such an RG structure, when present, makes the the tensor network representation even more efficient. We give some general evidence for the existence of the desired RG structure, but we do not rigorously establish it.

\subsection{Main assumption: existence of local thermal equilibrium}

Given a thermalizing system, there generally exists a finite time $\tau$ to reach local equilibrium. Correspondingly, away from classical critical points (for the critical case see \cite{RevModPhys.49.435}),
correlations decay beyond a correlation length scale $\xi$, so the notion of local thermal equilibrium is sensible within a derivative expansion. Provided the effective local temperature, effective local chemical potential, etc. vary slowly on the scale of the correlation length, a local equilibrium state should provide a good approximation to the state of the system. For example, the temperature should satisfy $\xi \| \nabla T \|/T \ll 1$ for the derivative expansion to be valid.

Hence, to a good approximation, we expect that the NESS of an interacting system should be described by a state of the form
\beq
\rho_{\text{NESS}} = \frac{\exp\left(- \sum_x \beta(x) \mathcal{E}(x) + ... \right)}{Z_{\text{NESS}}},
\eeq
where $\mathcal{E}(x)$ is the energy density, $\beta(x)$ is a position dependent inverse temperature, the ellipses denotes further local operators like conserved currents, and $Z_{\text{NESS}}$ denotes a non-equilibrium partition function.

Any density matrix of the form $\rho \sim \exp(-\sum_x O_x)$ has bounded mutual information between a spatial region $A$ and its complement $\bar{A}$ \cite{2008PhRvL.100g0502W}, $\CI(A:\bar{A}) \propto |\partial A|$ with the coefficient depending on the size of the operators $O_x$. For a thermal equilibrium state of a local Hamiltonian, the bound reads $\CI \leq \frac{J}{T} \frac{|\partial A|}{a^{d-1}}$ where $J$ is the energy scale of the microscopic Hamiltonian, and $a$ is the lattice spacing.

In fact, this bound, which diverges as $T \rightarrow 0$, is often quite loose. For
$(1+1)$-d CFTs the mutual information scales like $\CI \sim \log(1/(aT))$ which does diverge as $T \rightarrow 0$ but this divergence is much weaker than  a $1/T$ divergence \cite{Calabrese:2009qy}. For higher dimensional CFTs the mutual information remains finite, roughly
$\CI \sim |\partial A| a^{1-d}$, even at $T=0$. Fermi liquid mutual information also only diverges as $\log\left(E_F/T\right)$ as $T\rightarrow 0$ \cite{PhysRevB.86.045109}. These examples show that there is considerably less mutual information at small $T$ than would be expected from the upper bound \cite{2008PhRvL.100g0502W} to low $T$. See Appendix \ref{app:entropyboosted} for more details.

Following this discussion, our main assumption is that NESS in interacting systems are in local thermal equilibrium. As a consequence, entropies and correlations behave as they do in thermal equilibrium up to corrections which are controlled in a derivative expansion. In particular, the entropy of a region $A$ is assumed to have the form
\bea
\label{eq:entropyform}
S(A) &=& \int_A c_1(T(x),...) + \int_{\partial A} c_2(T(x),...) \nonumber \\
     &+& \sum_{i>2} \int_{\partial A} c_i(T(x),...) f_i( K, R,\nabla T,...)  \nonumber \\
     &+& \CO(\ell^d e^{-\ell/\xi})
\eea
where $R$ and $K$ are intrinsic and extrinsic curvatures of $\partial A$, $\ell$ is a linear size of $A$, and $\xi$ is the correlation length. The coefficients $c_i$ are functions of the local thermal data, e.g. temperature $T(x)$, chemical potential $\mu(x)$, etc., and the functions $f_i$ are combinations of curvatures and other vector fields, e.g. $\nabla T$. In the free fermion case, the exponentially small corrections in \eqref{eq:entropyform} are those we identified in Figures \ref{fig:mutualinformation} and \ref{fig:conditionalmutualinformation}.

The key point of \eqref{eq:entropyform} is that the extensive contributions from $\int_A$ and $\int_{\partial A}$ will cancel when considering differences of entropies of regions sharing parts of the same bulk and boundary. There may also be contributions to the entropy from sharp corners, but these will also cancel when taking differences of entropies provided the regions involved are all larger than the thermal correlation length. The mutual information and conditional mutual information are precisely such differences of entropies. \eqref{eq:entropyform} implies that the mutual information between a region and its complement obeys an area law and decays exponentially for distantly separated regions. In the case of the conditional mutual information, \eqref{eq:entropyform} implies that it is approximately zero for any three regions ($A$,$B$,$C$) such that $B$ separates $A$ from $C$.

The entropic statement \eqref{eq:entropyform} of local thermal equilibrium is a strong but reasonable assumption. In the free fermion model studied above, we showed explicitly that the mutual information and conditional mutual decay exponentially with distance. These results should generalize to any dimension and ultimately rely just on the decay of correlations. In a very different limit, one can also show, using AdS/CFT duality, that certain strongly interacting quantum field theories obey \eqref{eq:entropyform} \cite{2016arXiv160705753S}. Here the key insight is that entropies are computed using the Ryu-Takayanagi minimal area formula \cite{2006PhRvL..96r1602R}, and in a finite temperature black hole geometry where the minimal surfaces hug the black hole horizon, the Ryu-Takayanagi formula can be manipulated to give \eqref{eq:entropyform}. This remains true even for black holes in local thermal equilibrium, such as those describing a general hydrodynamic flow \cite{2008JHEP...02..045B}. A detailed demonstration of this will be reported elsewhere \cite{SwingleHubeny}.

Thus in the limit of weak and very strong interactions, the expansion \eqref{eq:entropyform} remains valid for a variety of non-equilibrium steady states in local thermal equilibrium. More generally, it is tied to the validity of a derivative expansion for the entropy of a general density matrix describing local thermal equilibrium. This derivative expansion should be sensible for any system which can be heated to infinite temperature without causing a phase transition.

\subsection{Tensor networks from approximate conditional independence}

In this subsection we demonstrate the existence of efficient tensor networks for NESS under the assumption of local thermal equilibrium. We focus on the basic intuition, the key assumptions, and the meaning of the results for transport calculation; details of the construction have been presented elsewhere \cite{2016arXiv160705753S}.

\begin{figure}
\includegraphics[width=.49\textwidth]{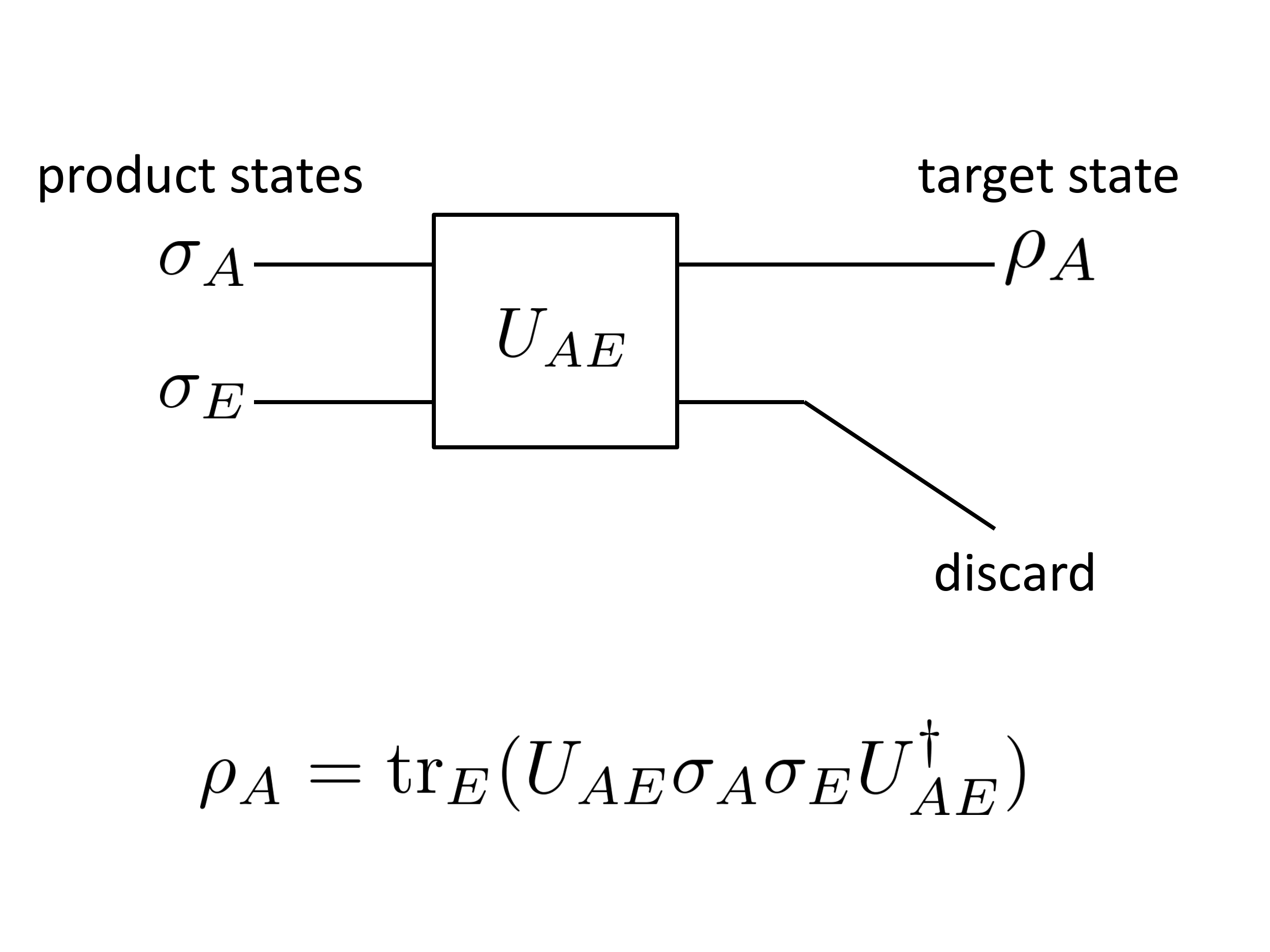}
\caption{Quantum channel which prepares the state $\rho_A$ from a product state on $AE$ after acting with a local unitary $U_{AE}$ and discarding $E$.}
\label{fig:channel}
\end{figure}

What we argue is that a state in local thermal equilibrium obeying \eqref{eq:entropyform} can be obtained as the output of a local quantum channel. In concrete terms this means that the state $\rho_A$ of interest must be obtainable as follows (see Figure \ref{fig:channel}):
\begin{enumerate}
\item Introduce a product state in an enlarged system $AE$. $AE$ must share the same geometry as $A$, i.e. we may add extra degrees of freedom on each site, but the locality properties of $A$ must not be modified.
\item Act with a local unitary transformation $U_{AE}$ on $AE$. Local is defined with respect to the metric on $A$ which $AE$ inherits.
\item Trace out $E$ to produce a state on $A$.
\end{enumerate}
The range of the local unitary $U_{AE}$ in step 2 is of order the correlation length.

To understand the construction we must introduce some notation. Given three regions $A$, $B$, and $C$, we say that $A$ is conditionally independent of $C$ given $B$ if the condition mutual information, $\CI(A:C|B)$, defined in \eqref{eq:cmidef} vanishes.

For classical random variables conditional independence occurs when the probability distribution satisfies $p(a,b,c) = p(a,b)p(b,c)/p(b)$. This condition implies that the full probability may be obtained from the marginals, that is $p(a,b,c)$ can be reconstructed from just $p(a,b)$ and $p(b,c)$. The reconstruction is just Bayes' rule:
\bea
p(a,b,c) &&=  p(c|b,a)p(a,b)  \nonumber\\
&&=_{\CI=0} p(c|b)p(a,b) = \frac{p(c,b)}{p(b)}p(a,b).
\eea

There is a quantum analogue of the classical reconstruction theorem, namely that if $\CI(A:C|B)=0$ then there exists a quantum channel $\CN_{B\rightarrow BC}$ such that $\text{Id}_A \otimes \CN_{B\rightarrow BC}(\rho_{AB}) = \rho_{ABC}$ \cite{petz1986,2003RvMaP..15...79P}. Moreover, recently it has been shown that this reconstruction theorem is approximately true provided $\CI$ is approximately zero \cite{fr1,fr2,2015RSPSA.47150338W,2015arXiv150907127J}. Our key observation is that with a suitable choice of geometry (meaning the regions $A$, $B$, and $C$), a NESS in local thermal equilibrium has $\CI \approx 0$, so an approximate reconstruction is possible.

Again, what this means in practical terms is that the thermal state can be obtained from a product state by introducing an environment, acting with a local unitary on the system plus environment, and then tracing out the environment. The actual construction proceeds by introducing a ``fattened" cellular decomposition of space. Here fattened means that all the cells have several correlation lengths thickness. Then if, for all $n$, the state of the $n$-cells is conditionally independent of the state of the $(n-1)$-cells given an interface, the thermal state is the output of a short-range quantum channel and has an efficient tensor network representation. Physically, we are patching together local data to form the global state.

As an example, consider a $(1+1)$-d CFT at finite temperature. The entropy of a region of length $\ell$ is
\beq
S(\ell,T) = \frac{c}{3} \log\(\frac{\sinh(\pi T \ell)}{\pi T a}\)
\eeq
where the velocity has been set to $1$ and $a$ is again the UV cutoff \cite{Calabrese:2009qy}. Now suppose $A = [0,\ell]$, $B=[\ell,2\ell]$, and $C=[2\ell,3\ell]$. The conditional mutual information is
\beq
\CI(A:C|B) = \frac{c}{3} \log\(\frac{\sinh^2(2\pi T\ell)}{\sinh(\pi T \ell) \sinh(3\pi T \ell)} \) \approx 0
\eeq
provided $\ell \gg 1/T$. This implies that the state of $ABC$ can be reconstructed from the state of $AB$ and $BC$ using a local channel of range $\ell$.

\begin{figure}
\includegraphics[width=.49\textwidth]{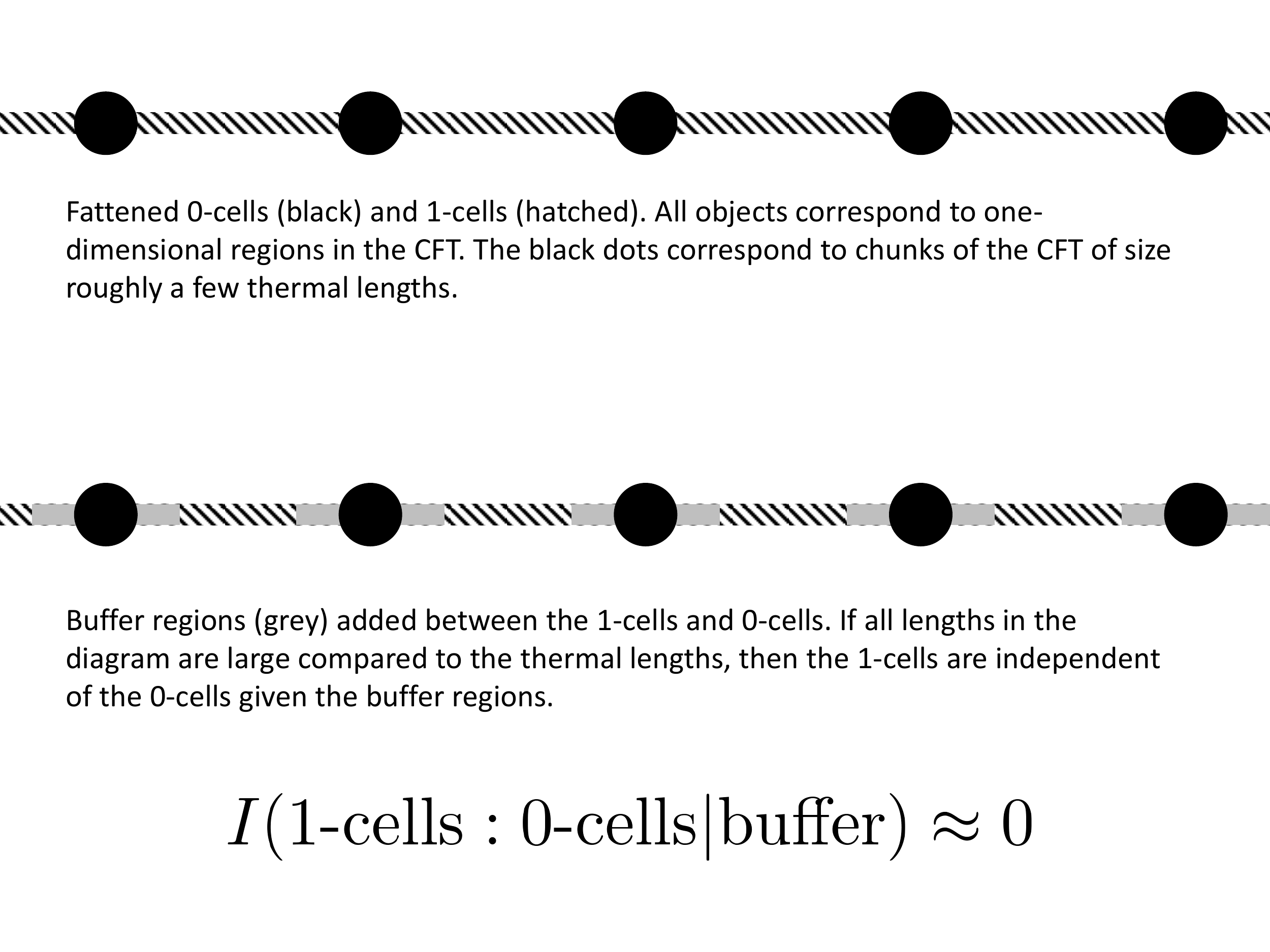}
%\caption{Setup of the approximate conditional independence construction in 1d.}
%\label{fig:1dmarkov}
%\end{figure}
%\begin{figure}
\includegraphics[width=.4\textwidth]{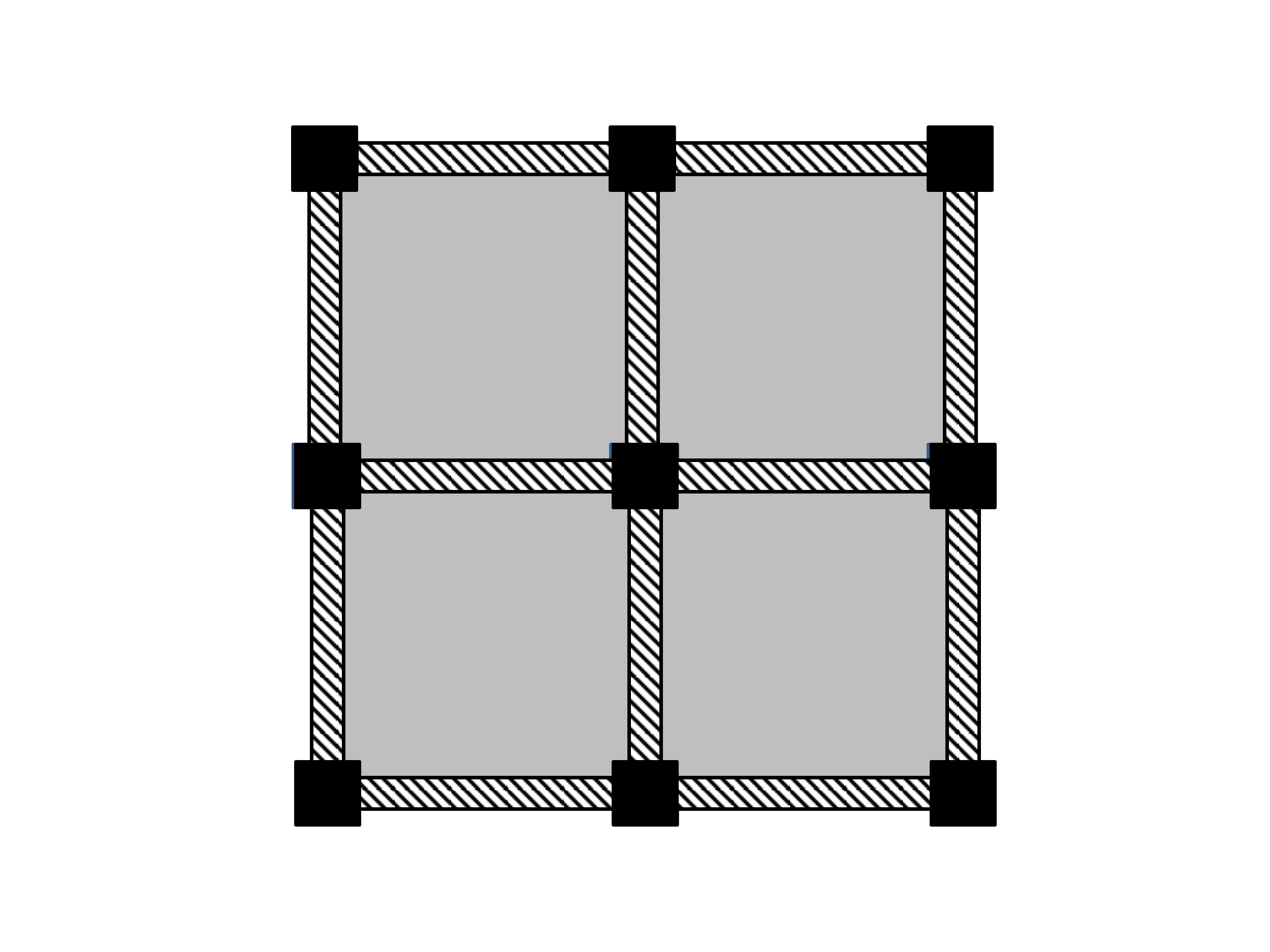}
\caption{Top: Setup of the approximate conditional independence construction in 1d. Bottom: Setup of the approximate conditional independence construction in 2d. The system is assembled starting from the black $0$ cells by adding the hatched $1$ cells and finally the grey $2$ cells.}
\label{fig:markov}
\end{figure}

To reconstruct all of space we decompose the line into a sequence of intervals as shown in the top part of Figure \ref{fig:markov}. Then if the state of the $1$ cells is independent of the state of the $0$ cells given an interface region, then there exists a quantum channel which approximately attaches the $1$-cells to the $0$-cells using only local operations. In fact, all the theorem guarantees is that there is a channel acting simultaneously on all the $1$-cells (but just the $1$-cells) which performs the reconstruction, but because the mutual information between different $1$-cells is approximately zero the reconstruction channel may be chosen to be local.

Of course, the argument also generalizes immediately to higher dimensions; see bottom part of Figure \ref{fig:markov} for the setup in 2d. As before, provided \eqref{eq:entropyform} is obeyed the relevant conditional mutual informations and mutual informations are approximately zero and approximate local reconstruction channels exist.

The picture again is that the state of the system is built from thermal scale pieces assembled using local operations. In the next subsection, we give a renormalization group (RG) argument which indicates that the thermal scale pieces may themselves be assembled hierarchically from strictly microscopically local pieces.

\subsection{Renormalization group structure of tensor networks}

The previous discussion guaranteed only that the state could be prepared using thermal scale operations. However, at low temperatures the thermal length scale is quite long compared to microscopic scales, and the general thermal scale result is not sufficient to guarantee a computationally efficient tensor network. However, in both the free fermion and holographic examples there is an additional renormalization group structure which permits to further decompose the thermal scale operations into a sequence of microscopic scale operations which are more efficient than a general thermal scale operation. We conjecture that this renormalization group structure generalizes beyond the free fermion and holographic examples and give some evidence for the conjecture.

Take first the case of free fermions in 1d. There it is known from numerical studies
\cite{2010PhRvB..81w5102E,2016PhRvL.116n0403E} that there exist approximate microscopic range unitary transformation which take the ground wavefunction on a size $L$ system into the ground state wavefunction on a size $L/2$ system times some extra product states. This unitary transformation can be interpreted as a kind of renormalization group transformation. Considering the ground state of a size $L$ system, the RG transformation can be repeated roughly $\log_2 L$ times before the system shrinks to microscopic size. Starting from the final factorized state and applying the unitaries in reverse then produces a $\log_2 L$ depth circuit which prepares the ground state.

The short-distance correlations in the ground state are not strongly modified by non-zero temperature. Thus, the general thermal scale unitary transformation $U_{AE}$ can be further decomposed into short-range pieces. The result approximates the ground state circuit cutoff at the thermal scale, plus some additional mixing between $A$ and $E$ at the thermal scale which produces the entropy of $\rho_A$.

Now consider the case of the strongly interacting holographic model. It has been argued that the holographic geometry may also be understood as being composed of a renormalization group circuit \cite{mera,2012PhRvD..86f5007S}. In so far as the black hole cuts off the holographic geometry deep inside the bulk while essentially preserving the geometry near the boundary, we may also speculate that a modified version of the ground state circuit with an infrared cutoff at the thermal scale can construct the thermal state from a product state. This is physically reasonable and progress has been made showing it explicitly \cite{2012PhRvD..86f5007S,2013PhRvD..87f6002M,2014GReGr..46.1823M,2015arXiv151007637C,2016arXiv160705753S}.

To recap the argument why we should be able to decompose $U_{AE}$ into simpler pieces, recall that even for a CFT ground state, the mutual information between a region and its complement is proportional to the area of the region, up to logarithmic corrections. By contrast, the statement that the thermal state can be produced using thermal scale operations acting on patches of linear size $\xi(T)$ would allow as much as $R^{d-1} \xi(T)$ mutual information for a region of linear size $R$. Since $\xi(T)$ diverges as the temperature approaches zero, this is far too much correlation compared to what we expect for the ground state.
The breakup of $U_{AE}$ into smaller pieces lowers the amount of correlation produced by the network and so brings it in line with thermal state expectations. Moreover, we have evidence at weak and strong coupling that such a decomposition exists.

\section{Discussion}

In this paper we have taken an entanglement-based perspective on the physics of quantum transport. Our main result is to establish the existence of tensor network representations of non-equilibrium steady states (NESS) which are in local equilibrium. We also argued that the NESS could be determined by evolving a suitable Markovian dynamics until a fixed point is reached. We illustrated this physics with a detailed study of free fermions and compared to known results for the fermion model. Based on these results, we proposed that open system dynamics within a manifold of lightly entangled states is a general framework for the calculation of transport properties of interacting quantum many-body systems.

Indeed, with the guarantee that the NESS has an efficient tensor network representation and assuming the relaxation rate of the Lindblad equation is polynomial in system size, then it follows that transport properties of even strongly interacting quantum many-body systems can be computed in time polynomial in system size --- a major improvement over exact diagonalization and closed system entanglement methods. Practically speaking, it is a non-trivial technical challenge to design the right law of dynamical evolution, but here we can appeal to the principle that electrical transport for macroscopic samples doesn't depend too much on the details of the contact. Working with tensor network states and Lindblad dynamics is also a non-trivial technical challenge, one which we address in a future paper. It would also be very interesting to combine our results with Keldysh methods, for example with the Meir-Wingreen formula \cite{PhysRevLett.68.2512}.

The systems where we believe the entanglement based approach has the greatest comparative advantage are those with strong correlations which lack a quasi-particle description. Especially difficult is the inclusion of quenched disorder in strongly interacting models, but the entanglement approach is expected to function even better with disorder. Indeed, disorder is known to reduce the average entanglement of a metal from area law violating to area law obeying \cite{2014arXiv1408.1094P}. Hence disordered strongly interacting non-Fermi liquids provide one ideal target for these methods. We are also interested in thermal transport in strongly coupled conformal field theories in two and three dimensions, say the quantum Ising model at its critical point. Another very tempting target is the infamous linear $T$ resistivity in cuprate superconductors.

In setting up these calculations, it is important to correctly incorporate conservation laws. Typically charge is conserved within the sample but can be exchanged with the environment at the boundary. Energy is more subtle; in the fermion examples we consider above energy was also conserved within the sample, but depending on the physical setting it will be more appropriate to allow energy to dissipate in the bulk, e.g. treating phonons as a bulk energy sink. Bulk energy dissipation is interesting since it is expected to further decrease the amount of entanglement present in the NESS and may speed up the approach to equilibrium.

For tensor network calculations there are two primary ingredients. First, we must dynamically evolve within the tensor network variational class. This can be accomplished by applying the Lindblad evolution using tensor network operators. The time evolved state is then truncated back into the low entanglement subspace, and our arguments above imply that this truncation shouldn't entail much error. Second, physical properties of the state must be computed. This is also a non-trivial task for general tensor networks, but the short-ranged nature of entanglement and correlations make this calculation feasible.

Finally, let us speculate briefly about the larger connections between tensor networks and quantum gravity.
It seems that holographic duality can be understood as a kind of mean field theory of tensor networks. It then becomes interesting to understand how to implement the dissipative dynamics of a black hole in the tensor network framework. Within the AdS/CFT duality, the conformal field theory is a closed quantum system with unitary dynamics, so the apparently dissipative black hole dynamics results from the transfer of quantum information from simple to complex degrees of freedom which have been coarse-grained. This information transfer is dual to the process of falling into the black hole, and it is this process which we would like to translate into tensor network language.

The use of open-system dynamics provides one physical way to implement a time evolution which effectively coarse-grains the system (see Appendix \ref{coarse-v-fine}), but it requires the explicit modeling of the contacts. A dynamical evolution law which uses only a system's own intrinsic tendency to transfer information into complex, inaccessible degrees of freedom is highly desirable. Such a dynamical law would in essence convert entanglement entropy to statistical entropy and would hide complexity the way a black hole does.

\acknowledgements
It is a pleasure to thank Cristian Zanoci for comments on the manuscript and John McGreevy for collaboration on closely related topics.
RM is supported by a Gerhard Casper Stanford Graduate Fellowship.
CDF is supported by the NSF Graduate Research Fellowship under Grant DGE-1106400.
NMT is supported through the Scientific Discovery through Advanced Computing (SciDAC) program funded by the U.S. Department of Energy, Office of Science, Advanced Scientific Computing Research and Basic Energy Sciences.
BGS is supported by the Simons Foundation and the Stanford Institute for Theoretical Physics.
We used the Extreme Science and Engineering Discovery Environment (XSEDE), which is supported by the National Science Foundation Grant No. OCI-1053575.

\appendix

\section{Derivation of Lindblad equation in Green's function form}
\label{fflindblad}

In this appendix we review how in free fermion systems the many-body Lindblad equation can be simplified to a single particle equation for the Green's function (essentially Wick's theorem). Consider a free fermion system with density matrix
\beq
\rho(t) = \frac{\exp{\left(- c^\dagger K(t) c\right)}}{Z(t)},
\eeq
where $K(t)$ is a time dependent single particle matrix.  Such a quadratic fermion state is totally determined by its two point function
\beq
G_{\alpha \beta}(t) = \text{tr}\left( c^\dagger_\alpha c_\beta \rho(t)\right).
\eeq
In fact, one easily finds that
\beq
G(t) = \frac{1}{e^{K^T(t)}+1}
\eeq
as matrices.  Note the transpose, a consequence of the matrix notation.  If we diagonalize $K$ and switch to the energy basis, then $G$ is the usual Fermi function.

Now suppose $\rho$ evolves in contact with an environment according to a master equation,
\beq
\partial_t \rho = -i [H,\rho] + \sum_j L_j \rho L_j^\dagger - \frac{1}{2} \sum_j \left\{ L_j^\dagger L_j , \rho \right\}.
\eeq
We take $H$ to be quadratic,
\beq
H = c^\dagger h c,
\eeq
and the jump operators $L_j$ are linear in the fermions, e.g.
\begin{align}
L_j &= u^\dagger_j c, \quad\text{or}\\
L_j &= c^\dagger v_j
\end{align}
with $u$ and $v$ column vectors of amplitudes.

Such a master equation preserves the Gaussian nature of the state, hence we should be able to understand the dynamics by studying just the equation of motion for $G$.  We compute first the Hamiltonian term.  We must evaluate
\bea
\left[\partial_t G\right]_{\text{cons}} &&= \tr\left(c^\dagger_x c_y [-i H,\rho]\right) \nonumber\\
&&= \tr\left([c^\dagger_x c_y ,-i H]\rho\right) \nonumber\\
&&= -i \sum_{z,w} h_{z,w} \tr\left([c^\dagger_x c_y ,c_z^\dagger c_w]\rho\right).
\eea
The commutator is
\beq
[c^\dagger_x c_y , c^\dagger_z c_w] = c^\dagger_x c_w \delta_{y,z} - c^\dagger_z c_y \delta_{x,w},
\eeq
and because we subsequently average over $\rho$, the final answer may be cast as
\beq
\left[\partial_t G\right]_{\text{cons}} = -i [G,h^T].
\eeq
As a check, if $K= h$, then because $[h^T,h^T]=0$, $G$ is time independent.  Again, note that the appearance of the transpose is just a consequence of the matrix notation and is of no deep significance.

Now we turn to the dissipative terms.  The general term in the sum over $j$ (sum over jump operators) is
\bea
&& \left[\partial_t G\right]_{\text{diss}} =  \tr\left(c^\dagger_x c_y L \rho L^\dagger  - c^\dagger_x c_y \frac{1}{2}\left\{L^\dagger L, \rho\right\}\right) \cr \nonumber
&& = \tr\left(\left(L^\dagger c^\dagger_x c_y L - \frac{1}{2}\left\{L^\dagger L, c^\dagger_x c_y\right\}\right)\rho\right).
\eea
If $L = u^\dagger c$ then we compute
\beq
c_a^\dagger c^\dagger_x c_y c_b - \frac{1}{2} \left\{c_a^\dagger c_b , c_x^\dagger c_y \right\} = -\frac{1}{2} \left( c_a^\dagger c_y \delta_{b,x} + c_x^\dagger c_b \delta_{a,y} \right).
\eeq
Appending the factors of $u$ and taking the average over $\rho$ we have
\bea
\left[\partial_t G\right]_{\text{diss}} &&=  \sum_{a,b}-\frac{1}{2} u^*_b u_a \left( G_{a,y} \delta_{b,x} + G_{x,b} \delta_{a,y} \right) \nonumber \\ &&= -\frac{1}{2} \left\{ u^* u^T, G \right\}.
\eea
If $L = c^\dagger v$ then we find
\bea
&& \left[\partial_t G\right]_{\text{diss}} = \frac{v_b v^*_a}{2} \tr\left(\left(c_a c_x^\dagger \delta_{y,b} + c_y c_b^\dagger \delta_{a,x} \right)\rho \right) \cr \nonumber
&& = v^* v^T - \frac{1}{2} \left\{v^* v^T,G \right\} \cr \nonumber
&& = \frac{1}{2} \left\{ v^* v^T, 1 - G\right\}.
\eea
If we now add up all terms we find
\beq \label{markovG}
\partial_t G = i [h^T,G] - \frac{1}{2}\sum_j \left\{u_j^* u_j^T,G\right\} + \frac{1}{2}\sum_j \left\{v_j^* v_j^T,1-G\right\}.
\eeq
This equation completely determines the two point function $G$ which in turn completely determines the state.  We can also directly compute the von Neumann entropy of regions using $G$.

\section{Lindblad operators for free fermion thermal states}
\label{ffthermallindblad}

In this appendix we review a construction of the Lindblad operators such that the corresponding Lindblad equation has as its unique fixed point a free fermion thermal state. We begin with a single mode: given a single fermion mode $c$ with Hamiltonian $H = \epsilon c^\dagger c$, what Lindblad operator has the thermal state at temperature $T$ and chemical potential $\mu$ as its fixed point?

There will be only two jump operators, $L_1 = \sqrt{\gamma_{in}} c^\dagger$ and $L_2 = \sqrt{\gamma_{out}} c$, so just the ratio of their rates must be determined. The Lindblad fixed point equation is
\beq
0 = -i[H,\rho] + L_1 \rho L_1 + L_2 \rho L_2 - \frac{1}{2} \{ L_1^\dagger L_1 + L_2^\dagger L_2,\rho\}.
\eeq
If $\rho = \exp(-\beta(H-\mu n))$ is to be a solution, then the Hamiltonian term drops out and only the jump operators contribute.

To find the right rates it helps to use the operator relation $f(n) c = c f(n-1)$ where $n=c^\dagger c$. The equation to satisfy is
\bea
\gamma_{in} c^\dagger e^{-\beta(\epsilon-\mu)n} c + &&\gamma_{out} c e^{-\beta(\epsilon-\mu)n} c^\dagger = \nonumber\\
&&(\gamma_{in} c c^\dagger + \gamma_{out} c^\dagger c) e^{-\beta(\epsilon-\mu)n}.
\eea
The correct ratio of rates is
\beq
\gamma_{in} e^{\beta(\epsilon-\mu)} = \gamma_{out}.
\eeq

Then for any multimode system (but still free), the same construction immediately generalizes to give a Lindblad operator whose fixed point is the thermal state. In more detail, let the single particle Hamiltonian be $h$ with energy levels $\epsilon_n$ and single particle wavefunctions $\psi_n$. The wavefunctions $\psi_n$ are functions of position $\psi_n(x)$ or in our case functions of a discrete position or lattice site label. For each mode $\epsilon_n$ we then have a pair of jump operators of the form $\sqrt{\gamma_{in,n}} c^\dagger_n$ and $\sqrt{\gamma_{out,n}} c_n$. Note that $\sum_x \psi^*_n(x) c(x) = c_n$ is just the annihilation operator for the energy mode $n$. As in the single mode case, the only constraint is that
\beq
\gamma_{in,n} e^{\beta (\epsilon_n -\mu)} = \gamma_{out,n}
\eeq
for all $n$.

Note that the correct change of basis in an $M$ mode system is easily remembered as follows. The Hamiltonian is $H = c^\dagger h c$. The eigenstates of $h$ are $h \psi_n = \epsilon_n \psi_n$ (remember that $h$ is just an $M \times M$ matrix, while $H$ is a quantum operator on the many-particle Hilbert space). Assemble the eigenvectors of $h$ into a matrix $u_h = [\psi_1 ,....,\psi_M]$ and energies into a diagonal matrix $d_h = (\epsilon_1,...,\epsilon_M)$. Then using $h u_h = u_h d_h$ it follows that
\beq
H = c^\dagger h c = c^\dagger u_h d_h u_h^\dagger c = (u_h^\dagger c)^\dagger d_h (u_h^\dagger c).
\eeq
Hence the correct modes are given by the entries of the vector of operators $u_h^\dagger c$.

\section{Solution of the two-site model}
\label{app:twosite}
The steady-state equation for the correlation matrix reads
\beq
0 = -i w [X,G] -\frac{1}{2}\{\gamma r_0 + \gamma \delta r Z,G \} - \frac{1}{2} \{\gamma ,G-1\}.
\eeq
To simplify notation we have switched to using the Pauli operators $\sigma^i \sim X,Y,Z$ and the identity $1$. The equation is most easily solved by resolving $G$ in the Pauli basis,
\beq
G=a_0 + a_y Y + a_z Z,
\eeq
with no $X$ term needed. Using $\{1,\sigma^i\} = 2 \sigma^i$, $\{\sigma^i,\sigma^j\} = 2\delta^{ij}$, and $[\sigma^i,\sigma^j] = 2i \epsilon^{ijk} \sigma^k$ the fixed point equation becomes
\bea
&&0 = -iw(2i a_y Z - 2ia_z Y) - \gamma(1+r_0) (a_0 + a_y Y + a_z Z) \nonumber\\ && - \gamma \delta r (a_0 Z + a_z) + \gamma.
\eea
Setting the coefficients of $1$, $Y$, and $Z$ equal to zero gives
\begin{align}
0 &= -\gamma(1+r_0) a_0 - \gamma \delta r a_z + \gamma,\\
0 &= -2 w a_z - \gamma(1+r_0)a_y,\\
0 &= 2 wa_y -\gamma (1+r_0)a_z - \gamma \delta r a_0.
\end{align}
Solving this set yields
\begin{align}
a_y &= -\frac{2w}{\gamma(1+r_0)} a_z,\\
a_0 &= \frac{1}{\gamma \delta r} \left( - \frac{4w^2}{\gamma(1+r_0)} - \gamma(1+r_0)\right) a_z,\\
a_z &= \delta r \left(-\frac{4 w^2}{\gamma^2} - (1+r_0)^2 +\delta r^2 \right)^{-1}.
\end{align}
Note that $a_z$ is negative if $\delta r $ is positive as is expected since then $\mu_L < \mu_R$ and hence $n_L < n_R$. It also follows that
\beq
a_y = \frac{2 w}{\gamma(1+r_0)} \frac{\delta r}{\frac{4 w^2}{\gamma^2} + (1+r_0)^2 - \delta r^2},
\eeq
which gives us the expression for the current.

\section{Review of 1d ballistic transport}
\label{sec:1dballistic}

Consider a chain of $L$ sites with Hamiltonian
\beq
H = \sum_r \left[ - w c_r^\dagger c_{r+1} + h.c. - \mu c_r^\dagger c_r \right].
\eeq
The Hamiltonian is easily diagonalized by going to the momentum basis. Writing
\beq
c_k = \sum_r \frac{e^{ikr}}{\sqrt{L}} c_r
\eeq
allows to express the Hamiltonian as
\beq
H = \sum_k (\epsilon_k - \mu) c_k^\dagger c_k,
\eeq
where $\epsilon_k = - 2 w \cos(k)$. We have taken units of length in which the lattice constant is one. The momentum $k$ is confined to the first Brillouin zone $k\in [-\pi,\pi)$. The state labelled by $k$ has a group velocity given by $v_k = \partial_k \epsilon_k = 2 w \sin(k)$.

Now the usual picture of transport through such a quantum wire is ballistic \cite{5392683,quantumtransportbook}. One imagines that right moving particles move through the wire coming from some distant left lead and similarly that left moving particles come from distant right lead. When there is no scattering in the wire, then the charge and energy currents are determined simply by the difference in the flows from the left and the right (this assumes that each particle moving from left to right or vice versa is instantly thermalized when it reaches the opposite lead.

This yields the following expression for the currents. The electric current (with the electron charge set to one) from left to right is given by
\beq
I_{L\rightarrow R} = \int_0^\pi \frac{dk}{2\pi} v_k f(\epsilon_k - \mu_L,T_L);
\eeq
similarly, the thermal current is
\beq
I_{E,L\rightarrow R} = \int_0^\pi \frac{dk}{2\pi} \epsilon_k v_k f(\epsilon_k - \mu_L,T_L).
\eeq
Here $f$ is assumed to be the Fermi function $f(E,T) = (e^{E/T}+1)^{-1}$ although in principle what appears here is simply the distribution function of the left lead (which we assume is thermal).
The corresponding currents from right to left are
\begin{align}
I_{R\rightarrow L} &= \int_{-\pi}^0 \frac{dk}{2\pi} v_k f(\epsilon_k - \mu_R,T_R),
\quad \text{and}\\
I_{E,R\rightarrow L} &= \int_{-\pi}^0 \frac{dk}{2\pi} \epsilon_k v_k f(\epsilon_k - \mu_R,T_R).
\end{align}
The total currents are
\begin{align}
I &= I_{L\rightarrow R}+I_{R\rightarrow L},\\
I_E &= I_{E,L\rightarrow R}+I_{E,R\rightarrow L}.
\end{align}

Let us first examine the limit of small $T$. In this case the electrical current reduces to the difference in the chemical potential between left and right. To show this we begin with the formula
\beq
\lim_{T\rightarrow 0} f(E,T) = \theta(-E)
\eeq
with $\theta(x)$ the Heaviside step function. Then it follows that
\beq
\int_0^\pi \frac{dk}{2\pi} \partial_k \epsilon_k \theta(\mu - \epsilon_k) = \int_0^\mu \frac{d\epsilon}{2\pi} = \frac{\mu}{2\pi}.
\eeq
Since the velocity has the opposite sign for $k \in (-\pi,0)$ it follows that the total current is
\beq
I = \frac{1}{2\pi} (\mu_L - \mu_R).
\eeq
A similar calculation reveals that to lowest non-vanishing order the energy current is ($\mu_L = \mu_R = 0$)
\beq
I_E = \frac{\pi}{12} (T_L^2 - T_R^2).
\eeq

What about the opposite limit of very high temperature, $T \gg w$? In this case one may expand the Fermi function as
\beq
f(E,T) = \frac{1}{2 + E/T} = \frac{1}{2} - \frac{E}{4 T} + ...
\eeq
The electrical current requires computing the integral
\beq
\int_0^\pi \frac{dk}{2\pi} v_k \left(\frac{1}{2} - \frac{\epsilon_k - \mu}{4T}\right).
\eeq
The first term will cancel when the $R\rightarrow L$ current is subtracted and the second term simplifies to
$\frac{w \mu}{2\pi T}$.
This gives an electrical current of
\beq
I = \frac{w}{2\pi T} (\mu_L - \mu_R).
\eeq

The energy current requires the integral
\beq
\int_0^\pi \frac{dk}{2\pi} \epsilon_k v_k \left(\frac{1}{2} - \frac{\epsilon_k - \mu}{4T}\right),
\eeq
where again the first term will cancel. Now, however, the $\mu$ term integrates to zero while the term involving $\epsilon_k$ from $f$ survives. The integral evaluates to
$-\frac{2 w^3}{3\pi T}$.
Note that there is a slight subtlety here about whether $\epsilon_k$ or $\epsilon_k -\mu$ is the energy being transported; we will discuss this later but one of the advantages of $\mu=0$ is that this issue is irrelevant.
The energy current then is
\beq
I_E = -\frac{2 w^3}{3\pi} \left(\frac{1}{T_L} - \frac{1}{T_R}\right).
\eeq
As a check, if $T_L > T_R$ then energy should flow from $L$ to $R$ and so $I_E > 0 $ since $1/T_L < 1/T_R$.

We can also ask about electric current with a scattering center in the wire, say a barrier of some height. In this case there will be some probability $\mathcal{T}(E)$ to transmit at energy $E$ as well as a probability $\mathcal{R}(E) = 1 - \mathcal{T}$ to reflect at energy $E$. The currents (at zero temperature and assuming no energy dependence in $\mathcal{T}$) are then
\begin{align}
I_{L\rightarrow R} &= \frac{\mu_L}{2\pi} \mathcal{T}, \quad \text{and}\\
I_{R\rightarrow L} &= - \frac{\mu_R}{2\pi}\mathcal{T}.
\end{align}
This expressions arise because the total current from the $L$ lead, say, is $\mu_L/2\pi$, but only a fraction $\mathcal{T}$ makes it through the scatterer with the other fraction being reflected back. The total current is
\beq
I = I_{L\rightarrow R} + I_{R \rightarrow L} = \frac{\mathcal{T}}{2\pi} (\mu_L - \mu_R).
\eeq
An analogous computation can be carried out for energy currents and for other approximations to $\mathcal{T}$ and choices of $T$.

Now as a final comment, all these computations apply for the infinite wire with the explicit assumption of left and right thermal equilibrium at infinity. Our model is explicitly a finite wire (which is more physical) which nevertheless should display much of this physics. One possibility for deviations comes from the contact between the lead and the wire which can lead to extra scattering and a contact resistance. Otherwise we expect the above formulas to be well borne out given a sufficiently large wire.

\section{Review of quantum Markov chains}
\label{app:qmarkov}

Here we review the physics of quantum Markov chains \cite{qmarkov}. We say a tripartite state $\rho_{ABC}$ forms a quantum Markov chain if the conditional mutual information $I(A:C|B) = S(AB)+S(BC)-S(B)-S(ABC)$ vanishes. Classically this would imply that the probability distribution factorizes, $p(a,b,c) = \frac{p(a,b)p(b,c)}{p(b)}$, so that $A$ is independent of $C$ given $B$: $p(a,c|b) =  \frac{p(a,b,c)}{p(b)} = \frac{p(a,b)}{p(b)} \frac{p(b,c)}{p(b)}$. What is the quantum version of this statement?

Notice that strong subadditivity implies that $I(A:C|B) \geq 0$, so states with $I(A:C|B)=0$ saturate strong subadditivity. Suppose $\rho_{ABC}$ saturates strong subadditivity and consider a generic perturbation $\rho_{ABC} \rightarrow \rho_{ABC} + \delta \rho_{ABC}$. Demanding that the trace of the perturbed state be $1$ gives $\text{tr}(\delta \rho_{ABC})=0$. Now consider the variation of $I(A:C|B)$ and use $\delta S(\sigma) = - \tr( \delta \sigma \log \sigma) - \tr(\delta \sigma) = - \tr(\delta \sigma \log \sigma)$. We find
\begin{align}
\delta I(A:C|B) = \tr(\delta \rho_{ABC} [&-\log \rho_{AB} - \log \rho_{BC} \nonumber \\
&+ \log \rho_B + \log \rho_{ABC}] ),
\end{align}
but $I(A:C|B)$ must remain positive for all $\delta \rho$ so the linear term in the variation must vanish,
\beq
\log \rho_{ABC} = \log \rho_{AB} + \log \rho_{BC} - \log \rho_B,
\eeq
which is the quantum analog of $p(a,b,c) = p(a,b)p(b,c)/p(b)$. Note: we have not been careful with the factors, but expressions like $-\log \rho_{BC}$ should be understood as $- I_A \otimes \log \rho_{BC}$ as appropriate.

What about reconstruction? In the classical case we can exactly reconstruct $p(a,b,c)$ from $p(a,b)$ and $p(b,c)$ if the conditional mutual information is zero. The reconstruction is Bayes rule,
\begin{align}
p(a,b,c) &= p(c|a,b) p(a,b) \nonumber \\
&\underbrace{\rightarrow}_{\CI=0} p(c|b) p(a,b) = \frac{p(b,c) p(a,b)}{p(b)}.
\end{align}
The quantum analog of this statement is Petz's recovery channel,
\beq
\CR(\sigma_{AB}) = \rho_{BC}^{1/2} \rho_B^{-1/2} \sigma_{AB} \rho_B^{-1/2} \rho_{BC}^{1/2},
\eeq
which obeys $\CR(\rho_{AB}) = \rho_{ABC}$. There is an analogous formula in the case of approximate condition independence \cite{2015RSPSA.47150338W,2015arXiv150907127J}. What is highly non-trivial here, and essential for the constructions in the main text, is that the Petz map doesn't depend on the state of the whole system but only on the state of the marginals.

The recent developments in this field were initiated by Fawzi and Renner \cite{fr1} who showed that if $I(A:C|B) \approx 0$ then there is an approximate recovery map which acts only on $B \rightarrow BC$. Furthermore, it was later shown \cite{fr2} that this map is independent of $\rho_A$ meaning not only does it not act on the $A$ system, the $B\rightarrow BC$ map is also independent of the state on $A$.

\section{Entropic structure of boosted free fermion states}
\label{app:entropyboosted}

To put the entanglement results in the free fermion model in context, we review the calculation of the mutual information and conditional mutual information using low energy field theory. Consider first the one-dimensional case. Suppose we have a piece of wire of length $L$ containing a spin-less fermion field $c(x)$ and suppose each end of the wire is coupled to a large fermion bath. The left $L$ and right $R$ baths have temperatures $T_{L}$, $T_R$ and chemical potentials $\mu_{L}$, $\mu_R$. The two point function of fermion operators within the wire is
\begin{align}
G(x,y) & = \langle c^\dagger(x) c(y)\rangle \nonumber \\
& = \int_0^\infty \frac{dk}{2\pi} e^{-i k(x-y)} f(E_k - \mu_L,T_L) \nonumber\\
&+ \int_{-\infty}^0 \frac{dk}{2\pi} e^{-i k(x-y)} f(E_k - \mu_R, T_R)
\end{align}
where $f(E,T) = 1/(e^{E/T}+1)$
is the Fermi function. If $\mu_L = \mu_R$ and $T_L = T_R$ then this describes a wire in equilibrium at temperature $T$ and chemical potential $\mu$.  More generally, the above two point function represents a fermion system with the right movers equilibrated with a left bath (from which they originated) and the left movers equilibrated with a right bath (from which they originated).

To illustrate the physics, suppose that $\mu_L = \mu_R$ but $T_L > T_R$. Linearize the energy about about $\mu$ and call the Fermi velocity $v_F$ (which may in general depend on energy and differ between left and right). The particle current is $I=0$ while the energy current is $I_E \propto T_L^2 - T_R^2$ (see Appendix \ref{sec:1dballistic}).

How much entanglement is present in this current carrying state? Conformal field theory supplies the answer in the long-wavelength low energy limit\cite{Calabrese:2009qy}. In terms of the left and right central charges $c_{L,R}$ and the left and right temperatures $T_{L,R}$ the entropy of an interval of length $L$ is
\begin{align}
S(L) &= \frac{c_L}{6}\log\left(\frac{\sinh\left(\frac{\pi T_L L}{v_F(\mu_L)}\right)}{\pi T_L a/v_F(\mu_L)}\right) + (L \leftrightarrow R)
%&& +\frac{c_R}{6}\log\left(\frac{\sinh\left(\frac{\pi T_R L}{v_F(\mu_R)}\right)}{\pi T_R a/v_F(\mu_R)}\right)
\end{align}
where $a$ is a UV cutoff. This formula allows for the possibility that the Fermi velocity depends on the chemical potential (as it will in general). It demonstrates that even if $\mu_L \neq \mu_R$ and $T_L \neq T_R$ the entanglement structure is qualitatively the same as the thermal state (or ground state) at the appropriate temperature.

Given these entropies the mutual information of the left half the wire, $x \in [0,L/2]$, with the right half of the wire, $x \in [L/2,L]$, can be computed. It is
\begin{align}
 &\CI([0,L/2],[L/2,L])   \nonumber\\
 &\approx \frac{c_L}{6}\log\(\frac{v_F(\mu_L)}{2 \pi T_L a} \) + \frac{c_R}{6}\log\(\frac{v_F(\mu_R)}{2 \pi T_R a} \),
\end{align}
up to exponentially small corrections. It is identical to ground state mutual information formula with the region size replaced by the thermal correlation length $\xi(T) =
v/(2\pi T)$.

What about in higher dimensions? The analog of the biased wire is a boosted Fermi surface, i.e. a current carrying metallic state. Solution of the Boltzmann equation shows that the state of a metal in a uniform electric field is the same as the equilibrium state but with all momenta shifted by $\vec{Q} = e \vec{E} \tau$ with $e$ the particle charge, $\vec{E}$ the electric field, and $\tau$ the relaxation time. Typically this scattering time comes from phonons or impurities, both of which will modify the state of the system in a non-trivial way, but for the purpose of the example we abstract away this extra complication and study just the boosted state.

Two states related by such a momentum shift are connected by a simple unitary transformation given by
\beq
U_Q = \exp\left(i \sum_j Q_j \int d^d x\, x^j c^\dagger(x) c(x) \right).
\eeq
Observe, however, that $U_Q$ is actually a product of local unitaries,
%\beq
%U_Q = \prod_x \exp\left(i \sum_j Q_j  x^j c^\dagger(x) c(x)\right),
%\eeq
so application of $U_Q$ to a state does not change the entanglement properties of the state. This establishes that all entropies of the boosted state are the same as they were in the equilibrium state.

In particular, the mutual information between a region $A$ and its complement $\bar{A}$ is given by the mutual information in the equilibrium state. This equilibrium mutual information is given \cite{PhysRevB.86.045109} in $d$ spatial dimensions by
\beq
\CI(A,\bar{A}) \sim k_F^{d-1} |\partial A| \log\(\frac{E_F}{T}\).
\eeq
Hence in any dimension current carrying metallic states are lightly entangled in the sense of having area law mutual information.

\section{Coarse-grained versus fine-grained dynamics}
\label{coarse-v-fine}

Here we briefly comment on the important distinction between closed and open system dynamics and the physical setup in which entanglement ideas are most useful.

Consider a closed quantum many-body system initialized in some out of equilibrium state. On general grounds we expect that such a time evolving closed system will rapidly generate entanglement between its spatial parts. Even if the initial state is lightly entangled, say obeying an area law, regions which are not too large (compared to the whole system) will have volume law entanglement at late times. For concreteness, consider an atomic gas and suppose we are interested in the diffusion constant of the gas.

Assuming the closed system thermalizes as far as macroscopic observables are concerned, it is possible to extract the diffusion constant as the gas relaxes to macroscopic equilibrium. However, the presence of a volume law for entanglement entropy means that it would not be possible to simulate the time evolution using entanglement based methods, at least not for very long. A hydrodynamic simulation would be possible, but if our goal is to calculate the diffusion constant by following the dynamics then hydrodynamic simulations will not suffice since the diffusion constant must be input by hand.

However, suppose the same diffusion constant entering the closed system dynamics also governs diffusion in an open system configuration. As an example of an open system configuration, suppose we permit a density gradient between two well separated environmental contacts. Except for dissipation at the contacts, the dynamics is unitary, and the dynamics of the environment-system composite is presumably still unitary. What this means is that large amounts of entanglement will still be generated, but we expect, due to the irreversibility of the environment, that the system is predominantly entangled with the environment rather than with itself.

Hence if we are only interested in the state of the system and not the state of the environment, it follows from the monogamy of entanglement that the system is not highly entangled amongst its own parts. Nevertheless, we can compute the current between the two contacts using just the state of the system, so the physical information about the diffusion constant is preserved despite the coupling to the environment. We may further imagine beginning the system in equilibrium and slowly biasing the contacts with the environment. If this biasing is done adiabatically then the system will remain close to its non-equilibrium steady state at all times and hence the dynamics can be restricted to the low entanglement states we identified above.

The environment thus acts as a sink for entanglement which preserves the macroscopic transport physics of interest. To recapitulate, assuming the diffusion constant can be measured using both a closed and an open system setup, it makes sense to use entanglement methods to compute the diffusion constant in an open system setup.

\bibliography{refs}

\end{document}